\documentclass[aps, prd, letterpaper, 12pt, amsmath, amssymb, amsfonts, floatfix, nofootinbib, showkeys, showpacs,  superscriptaddress]{revtex4-1}
\usepackage[utf8]{inputenc}

\usepackage{amsmath,amssymb}
\usepackage{slashed}
\usepackage{siunitx}
\usepackage{bm}
\usepackage{color}
\usepackage{graphicx} 
\usepackage[colorlinks=true,linkcolor=blue,citecolor=blue]{hyperref}
\usepackage{feynmp-auto}

\usepackage{natbib}
\usepackage{cleveref}

\newcommand{\du}{\mathrm{d}}
\newcommand{\dd}{\,\du} 
\newcommand{\bb}[1]{\bm{\mathrm{#1}}} 
\newcommand{\abs}[1]{\left|#1\right|}
\newcommand{\tr}{\operatorname{tr}}
\newcommand{\im}{\operatorname{Im}}
\newcommand{\re}{\operatorname{Re}}

\newcommand\BR{\operatorname{BR}}
\newcommand\TRH{T_{\mathrm{rh}}}
\newcommand\TFO{T_{\mathrm{FO}}}
\newcommand\TSN{T_{\mathrm{SN}}}
\newcommand\mpl{m_{\mathrm{Pl}}}
\newcommand\Lchi{\Lambda_{\mathrm{chiPT}}}
\newcommand\Ldd{\Lambda_{dd}}
\newcommand\Lsd{\Lambda_{sd}}
\newcommand\Lnn{\Lambda_{\nu\nu}}
\newcommand\LNP{\Lambda_{\mathrm{NP}}}
\newcommand\gq[2]{g_{#1}^{#2}}
\newcommand\gqc[2]{g_{#1}^{#2*}}
\newcommand\gqp[2]{\tilde{g}_{#1}^{#2}}
\newcommand\gqpc[2]{\tilde{g}_{#1}^{#2*}}

\begin{document}
\unitlength=1.3mm

\title{Cosmological implications of the KOTO excess}

\author{Wolfgang~Altmannshofer}
\email{waltmann@ucsc.edu}
\author{Benjamin~V.~Lehmann}
\email{blehmann@ucsc.edu}
\author{Stefano~Profumo}
\email{profumo@ucsc.edu}
\affiliation{Department of Physics, University of California Santa Cruz, 1156 High St., Santa Cruz, CA 95064, USA and \\
Santa Cruz Institute for Particle Physics, 1156 High St., Santa Cruz, CA 95064, USA}


\begin{abstract}
The KOTO experiment has reported an excess of $K_L\to\pi^0\nu\bar\nu$ events above the Standard Model prediction, in tension with the Grossman--Nir (GN) bound. The GN bound heavily constrains new physics interpretations of an excess in this channel, but another possibility is that the observed events originate from a different process entirely: a decay of the form $K_L\to\pi^0X$, where $X$ denotes one or more new invisible species. We introduce a class of models to study this scenario with two light scalars playing the role of $X$, and we examine the possibility that the lighter of the two new states may also account for cosmological dark matter (DM). We show that this species can be produced thermally in the presence of additional interactions apart from those needed to account for the KOTO excess. Conversely, in the minimal version of the model, DM must be produced nonthermally. In this case, avoiding overproduction imposes constraints on the structure of the low-energy theory. Moreover, this requirement carries significant implications for the scale of reheating in the early Universe, generically preferring a low but observationally permitted reheating temperature of $\mathcal O(\SI{10}{\mega\electronvolt})$. We discuss astrophysical and terrestrial signatures that will allow further tests of this paradigm in the coming years.
\end{abstract}

\maketitle
\setcounter{tocdepth}{1}
\tableofcontents

\clearpage
\section{Introduction}

The rare kaon decays $K^+ \to \pi^+ \nu\bar\nu$ and $K_L \to \pi^0 \nu \bar \nu$ are widely recognized as very sensitive probes of new physics (NP). In the Standard Model (SM), the branching ratios of these decays are strongly suppressed, and can be precisely predicted~\cite{Brod:2010hi,Buras:2015qea} to be
\begin{align}
    \BR(K^+ \to \pi^+ \nu\bar\nu)_\text{SM} &=
        (8.4 \pm 1.0) \times 10^{-11} ~, \\
    \BR(K_L \to \pi^0 \nu\bar\nu)_\text{SM} &=
        (3.4 \pm 0.6) \times 10^{-11} ~.
\end{align}
On the experimental side, several $K^+ \to \pi^+ \nu\bar\nu$ candidate events have been observed by the E787/E949 experiment~\cite{Adler:2002hy,Anisimovsky:2004hr,Artamonov:2008qb} and the NA62 experiment~\cite{CortinaGil:2018fkc}, but a discovery of $K^+ \to \pi^+ \nu\bar\nu$ has still to be established. The current best limit on the branching ratio is from a preliminary analysis of NA62 data and reads~\cite{NA62talk}
\begin{equation} \label{eq:NA62}
    \BR(K^+ \to \pi^+ \nu\bar\nu)_\text{exp} < 2.44 \times 10^{-10}
    \quad \text{(95\% C.L.)}, 
\end{equation}
not far above the SM prediction.
The NA62 experiment aims to measure the SM branching ratio with $\mathcal O(10\%)$ uncertainty.
In the case of $K_L \to \pi^0 \nu \bar \nu$, the current most stringent bound on the branching ratio comes from the KOTO experiment~\cite{Ahn:2018mvc}, and is still 2 orders of magnitude above the SM prediction:
\begin{equation} \label{eq:KOTOlimit}
    \BR(K_L \to \pi^0 \nu\bar\nu)_\text{exp} <
        3.0 \times 10^{-9} \quad \text{(90\% C.L.)}.
\end{equation}
Interestingly, in the latest status update by KOTO~\cite{KOTO}, 4 events are seen in the signal box, with an expected number of $0.05 \pm 0.01$ SM $K_L \to \pi^0 \nu\bar\nu$ events and $0.05 \pm 0.02$ background events. One of the events has been identified as likely background. If the remaining events are interpreted as signal, one finds a branching ratio of $\BR(K_L \to \pi^0 \nu\bar\nu) \sim 2\times 10^{-9}$~\cite{Kitahara:2019lws}. A branching ratio of this size would be a spectacular discovery. Not only does it imply NP, it also violates the Grossman-Nir (GN) bound~\cite{Grossman:1997sk}, $\BR(K_L \to \pi^0 \nu\bar\nu) \lesssim 4.3 \times \BR(K^+ \to \pi^+ \nu\bar\nu) \lesssim 10^{-9}$ when combined with the NA62 constraint in \cref{eq:NA62}.
The GN bound is very robust in models where the $K \to \pi \nu\bar\nu$ decays are modified by heavy new physics well above the kaon mass. However, in the presence of light new physics, the GN bound can be violated and the observed events at KOTO may find an explanation~\cite{Fuyuto:2014cya,Kitahara:2019lws,Egana-Ugrinovic:2019wzj,Dev:2019hho,Jho:2020jsa,Liu:2020qgx,He:2020jzn,Ziegler:2020ize,Liao:2020boe,Gori:2020xvq,Hostert:2020gou,Datta:2020auq,Dutta:2020scq,Cline:2020mdt}.

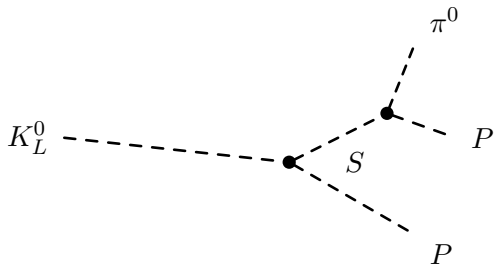
\begin{figure}
    \vspace{1cm}
    \begin{fmffile}{koto-signal}
        \begin{fmfgraph*}(40,20)
            \fmfleft{i1}
            \fmfright{o1,o2,o3}
            \fmf{dashes}{i1,v1}
            \fmf{dashes,label=$S$}{v1,v2}
            \fmf{dashes}{v1,o1}
            \fmf{dashes}{v2,o2}
            \fmf{dashes}{v2,o3}
            \fmfdot{v1,v2}
            \fmflabel{$K_L$}{i1}
            \fmflabel{$P$}{o1}
            \fmflabel{$P$}{o2}
            \fmflabel{$\pi^0$}{o3}
        \end{fmfgraph*}
    \end{fmffile}
    \caption{Decay chain accounting for the KOTO signal in our scenario.}
    \label{fig:koto}
\end{figure}

Here we focus on a new physics scenario first discussed in~\cite{Maxim_talk}. Two new light scalars $S$ and $P$, neutral under the SM gauge interactions, are introduced such that $K_L$ can decay into a pair of the new particles, $K_L \to SP$. If the decay $S \to \pi^0 P$ is allowed and $P$ is stable on the relevant experimental scales, then the decay chain $K_L \to SP \to \pi^0 PP$ can mimic the $K_L \to \pi^0 \nu\bar\nu$ signature (see \cref{fig:koto}). The corresponding chain of two-body decays does not exist for the charged kaon. A possible decay $K^+ \to \pi^+ SP$ is suppressed by three-body phase space or may be forbidden entirely by kinematics.

If $P$ is absolutely stable, it is also a candidate for cosmological dark matter (DM). In the minimal setup that can provide a NP explanation of the KOTO events, $P$ couples to the SM very weakly, implying that annihilation cross sections into SM states are too small for production by freeze-out. We therefore investigate alternative scenarios for cosmological production, and interpret overproduction of $P$ as a cosmological constraint on the structure of the low-energy theory. We show that $P$ is readily produced nonthermally if the scale of reheating is low, close to but safely above the current observational bound. We also show that this class of models can account for the KOTO excess without requiring a low reheating temperature, but only in the presence of additional interactions. We investigate prospects for testing this model with future experiments and with additional data from KOTO, and show that much of the parameter space will be probed in the near future.

This paper is organized as follows: In \cref{sec:model}, we present the model and discuss how it can explain the KOTO events. In \cref{sec:constraints}, we evaluate astrophysical and terrestrial constraints on the parameter space of our model. In \cref{sec:cosmological-production}, we consider cosmological production of $P$, and relate the production of $P$ to the scale of reheating. We discuss the implications of our results in \cref{sec:discussion} and conclude in \cref{sec:conclusions}.

\section{Model} \label{sec:model}
\begin{figure}[tb]
\centering
\includegraphics[width=.5\textwidth]{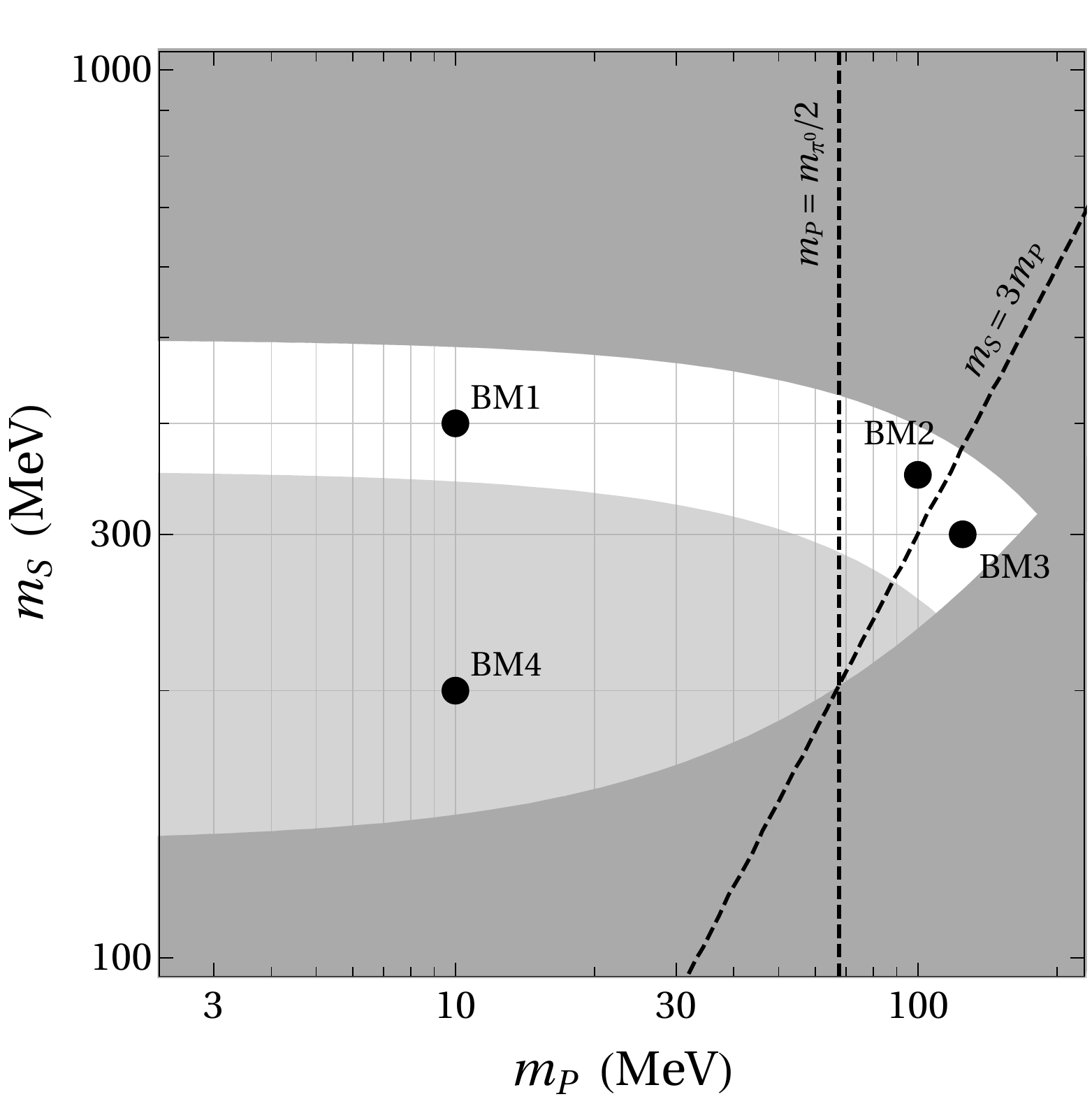}
\caption{The plane of the scalar masses $m_S$ vs. $m_P$. In the dark gray region the $K_L \to \pi^0 PP$ decay cannot be realized as a sequence of two-body decays.
In the light gray region the $K^+ \to \pi^+ SP$ decay is open. The black dots indicate four benchmark scenarios that we consider later (\cref{eq:benchmarks}).}
\label{fig:masses}
\end{figure}

We start with very simple kinematical considerations concerning the masses of the two scalars $S$ and $P$.
\Cref{fig:masses} shows the plane of the two scalar masses $m_S$ and $m_P$. As described in the Introduction, we are interested in regions of parameter space where the decay $K_L \to \pi^0 PP$, which mimics $K_L \to \pi^0 \nu\bar\nu$, can be realized as a sequence of the two-body decay $K_L \to SP$ followed by $S \to \pi^0 P$. For $m_S$ too large, the decay $K_L \to SP$ is kinematically forbidden, while for $m_S$ too small, the $S \to \pi^0 P$ decay is not open, excluding the dark gray regions in the plot. In the light gray region, one faces potential constraints from the charged kaon decay $K^+ \to \pi^+ SP$ that is generically expected in the models discussed below. In the white region, however, this decay is kinematically forbidden, while $K_L\to\pi^0\nu\bar\nu$ remains open.

The plot also indicates two other interesting kinematical boundaries. If $m_P < m_{\pi^0}/2$, the exotic pion decay $\pi^0 \to PP$ is possible which, as we will discuss in \cref{sec:cosmological-production}, can impact cosmological production considerably. If $m_S > 3m_P$, the decay $S \to 3P$ can be allowed, thus modifying the lifetime of $S$, which is a crucial parameter for beam-dump constraints. Note that low $P$ masses may be subject to constraints from supernova cooling, which we will discuss further in \cref{sec:supernova}. A weaker lower bound on the $P$ mass also follows from assuming a particular thermal history, a point to which we shall return in \cref{sec:discussion}.

In the following sections, we will discuss 4 benchmark parameter points covering the most interesting regimes:
\begin{equation}\label{eq:benchmarks}
    \begin{array}{ll}
    \text{BM1:}& \quad m_S = \SI{400}{\mega\electronvolt},
                 \quad m_P =  \SI{10}{\mega\electronvolt}, \\
    \text{BM2:}& \quad m_S = \SI{350}{\mega\electronvolt},
                 \quad m_P = \SI{100}{\mega\electronvolt}, \\
    \text{BM3:}& \quad m_S = \SI{300}{\mega\electronvolt},
                 \quad m_P = \SI{125}{\mega\electronvolt}, \\
    \text{BM4:}& \quad m_S = \SI{200}{\mega\electronvolt},
                 \quad m_P =  \SI{10}{\mega\electronvolt}.
    \end{array}
\end{equation}

Next we discuss in detail the interactions of $S$ and $P$ with SM quarks. We first focus on nonrenormalizable effective couplings and identify viable regions of parameter space. Then we comment on simplified UV models that map onto the effective couplings.

\subsection{Effective interactions of the scalars and meson decay rates}
\label{sec:BRs}

We assume that the scalars $S$ and $P$ interact with SM particles via the effective couplings
\begin{multline}\label{eq:Leff}
    \mathcal L_\text{int} \supset
    iSP \left(
        \frac{\gq{dd}{SP}}{\LNP} (\bar dd) +
        \frac{\gqp{dd}{SP}}{\LNP} (\bar d i \gamma_5 d) +
        \frac{\gq{ss}{SP}}{\LNP} (\bar ss) +
        \frac{\gqp{ss}{SP}}{\LNP} (\bar s i \gamma_5 s)
    \right) \\ 
    + iSP \left(
        \frac{\gq{sd}{SP}}{\LNP} (\bar sd) +
        \frac{\gqp{sd}{SP}}{\LNP} (\bar s i \gamma_5 d) +
        \mathrm{H.c.}
    \right).
\end{multline}
The factors of $i$ in the above Lagrangian are reminiscent of considering $S$ to be a CP-even scalar and $P$ to be a CP-odd pseudoscalar, a notational pattern that we will retain when matching onto low-energy QCD later on.
The coefficients $\gq{dd}{SP}$, $\gq{ss}{SP}$, $\gqp{dd}{SP}$, and $\gqp{ss}{SP}$ are purely imaginary (by Hermiticity of the Lagrangian) while the $\gq{sd}{SP}$ and $\gqp{sd}{SP}$ coefficients can have an arbitrary complex phase. 
There could also be interactions involving $b$ quarks, but as long as they are not considerably larger than the interactions with the light quarks, their impact on phenomenology will be negligible. 

In the following, we will also entertain the possibility of additional interactions involving $P^2$ and $S^2$, of the form
\begin{multline} \label{eq:Leff2}
    \mathcal L_\text{int} \supset
    P^2 \left(
        \frac{\gq{dd}{P^2}}{\LNP} (\bar dd) +
        \frac{\gqp{dd}{P^2}}{\LNP} (\bar d i \gamma_5 d) +
        \frac{\gq{ss}{P^2}}{\LNP} (\bar ss) +
        \frac{\gqp{ss}{P^2}}{\LNP} (\bar s i \gamma_5 s)
    \right) \\
    + P^2 \left(
        \frac{\gq{sd}{P^2}}{\LNP} (\bar sd) +
        \frac{\gqp{sd}{P^2}}{\LNP} (\bar s i \gamma_5 d) +
        \text{H.c.}
    \right).
\end{multline}
While the interactions in \cref{eq:Leff2} are not directly relevant for the KOTO signal, they do have important implications for other meson decays and in particular for the DM phenomenology as we will discuss in \cref{sec:cosmological-production} below.

The decays relevant for an enhanced KOTO signal, $K_L \to SP$ and $S \to \pi^0 P$, are induced by the couplings $\re(\gqp{sd}{SP})$ and $\im(\gqp{dd}{SP})$, respectively. 
For the corresponding decay rates, we find
\begin{align}
    \label{eq:KLSP}
    \Gamma(K_L \to S P) &=
        \frac{1}{8\pi} \frac{f_K^2 m_{K_L}^3}{m_s^2} \left(
            \frac{\re(\gqp{sd}{SP})}{\LNP}
        \right)^2 \eta_\text{QCD}~\sqrt{
            \lambda\left(1,m_S^2/m_{K_L}^2,m_P^2/m_{K_L}^2\right)
        }, \\
    \label{eq:SpiP}
    \Gamma(S \to \pi^0 P) &=
        \frac{1}{128\pi} \frac{f_\pi^2 m_{\pi^0}^4}{m_S m_d^2} \left(
            \frac{\im(\gqp{dd}{SP})}{\LNP}
        \right)^2 \eta_\text{QCD}~\sqrt{
            \lambda\left(1,m_{\pi^0}^2/m_S^2,m_P^2/m_S^2\right)
        },
\end{align}
with the phase space function $\lambda(a,b,c) = a^2+b^2+c^2-2(ab+ac+bc)$.    
The down and strange quark masses in the above expressions should be interpreted as the $\overline{\text{MS}}$ masses at a renormalization scale of $\mu = 2$~GeV. Leading-log QCD corrections are then taken into account through the factor $\eta_\text{QCD}$,
\begin{align} \label{eq:QCDrunning}
    \eta_\text{QCD} &= \left(
        \frac{\alpha_s(m_t)}{\alpha_s(M)}
    \right)^{8/7} \left(
        \frac{\alpha_s(m_b)}{\alpha_s(m_t)}
    \right)^{24/23} \left(
        \frac{\alpha_s(2~\text{GeV})}{\alpha_s(m_b)}
    \right)^{24/25}, 
\end{align}
where $M$ is the scale of new physics that is responsible for the effective interactions of $S$ and $P$ with the SM quarks. Because of $SU(2)_L$ invariance, we expect $M \sim \sqrt{\LNP v}$, where $v= 246$~GeV is the vacuum expectation value of the SM Higgs.
Note that including the $\eta_\text{QCD}$ factor is equivalent to evaluating the down and strange masses in \cref{eq:KLSP,eq:SpiP} at the scale $M$.

The coupling $|\gq{sd}{SP}|$ can lead to the decay $K^+ \to \pi^+ SP$, if kinematically allowed.
The differential three-body decay rate of $K^+ \to \pi^+ SP$ is given by
\begin{multline} \label{eq:KpiSP}
    \frac{d\Gamma(K^+ \to \pi^+ SP)}{dq^2} =
        \frac{1}{256\pi^3} \frac{m_{K^+}^3}{m_s^2} \left(
            \frac{|\gq{sd}{SP}|}{\LNP}
        \right)^2 \eta_\text{QCD} \left(
            1-\frac{m_{\pi^+}^2}{m_{K^+}^2}
        \right)^2 \\
        \times \sqrt{\lambda\left(1,m_S^2/q^2,m_P^2/q^2\right)}
        \sqrt{\lambda\left(1,m_{\pi^+}^2/m_{K^+}^2,q^2/m_{K^+}^2\right)},
\end{multline}
where we estimated the relevant scalar form factor as $\langle \pi^+ |\bar s d| K^+\rangle \simeq (m_{K^+}^2 - m_{\pi^+}^2)/m_s$ and $q^2$ is the invariant mass of the $SP$ system, with $(m_P + m_S)^2 < q^2 < (m_{K^+} - m_{\pi^+})^2$. 

Similar to the $K_L \to SP$ decay, the interactions in \cref{eq:Leff} also lead to the exotic eta decay $\eta \to SP$, which has been identified as a possible source of the scalar $S$ at beam-dump experiments~\cite{Hostert:2020gou}. Neglecting $\eta$--$\eta^\prime$ mixing, we find
\begin{equation} \label{eq:etaSP}
    \Gamma(\eta \to S P) = \frac{3}{512\pi} \frac{f_\eta^2 m_\eta^3}{m_s^2}
    \left(
        \frac{2\im(\gqp{ss}{SP})-\im(\gqp{dd}{SP})}{\LNP}
    \right)^2 \eta_\text{QCD}~\sqrt{
        \lambda\left(1,m_S^2/m_\eta^2,m_P^2/m_\eta^2\right)
    }.
\end{equation}
For completeness, we also provide the expression for the decay $K_S \to SP$:
\begin{equation} \label{eq:KSSP}
    \Gamma(K_S \to S P) = \frac{1}{32\pi} \frac{f_K^2 m_{K_S}^3}{m_s^2} \left(
        \frac{\im(\gqp{sd}{SP})}{\LNP}
    \right)^2 \eta_\text{QCD}~\sqrt{
        \lambda\left(1,m_S^2/m_{K_S}^2,m_P^2/m_{K_S}^2\right)
    }.
\end{equation}

In the presence of the $P^2$ interactions in \cref{eq:Leff2}, there are additional exotic meson decays, $\pi^0 \to PP$, $\eta \to PP$, $K_{L/S} \to PP$, and $K^+ \to \pi^+ PP$, with the following decay rates:
\begin{align}
    \label{eq:piPP}
    &\Gamma(\pi^0 \to PP) =
        \frac{1}{64\pi} \frac{f_\pi^2 m_{\pi^0}^3}{m_d^2} \left(
            \frac{\re(\gqp{dd}{P^2})}{\LNP}
        \right)^2 \eta_\text{QCD}~\sqrt{1 - \frac{4 m_P^2}{m_{\pi^0}^2}}~,
    \\
    \label{eq:etaPP}
    &\Gamma(\eta \to PP) =
        \frac{3}{256\pi} \frac{f_\eta^2 m_\eta^3}{m_s^2} \left(
            \frac{2\re(\gqp{ss}{P^2})-\re(\gqp{dd}{P^2})}{\LNP}
        \right)^2 \eta_\text{QCD}~\sqrt{1 - \frac{4 m_P^2}{m_\eta^2}}~,
    \\
    &\Gamma(K_L \to PP) =
        \frac{1}{4\pi} \frac{f_K^2 m_{K_L}^3}{m_s^2} \left(
            \frac{\im(\gqp{sd}{P^2})}{\LNP}
        \right)^2 \eta_\text{QCD}~\sqrt{1 - \frac{4 m_P^2}{m_{K_L}^2}}~,
    \\
    \label{eq:KSPP}
    &\Gamma(K_S \to PP) =
        \frac{1}{4\pi} \frac{f_K^2 m_{K_S}^3}{m_s^2} \left(
            \frac{\re(\gqp{sd}{P^2})}{\LNP}
        \right)^2 \eta_\text{QCD}~\sqrt{1 - \frac{4 m_P^2}{m_{K_S}^2}}~,
    \\
    \label{eq:KpiPP}
    &\frac{\du\Gamma(K^+ \to \pi^+ PP)}{\du q^2} =
        \frac{1}{128\pi^3} \frac{m_{K^+}^3}{m_s^2} \left(
            \frac{|\gq{sd}{P^2}|}{\LNP}
        \right)^2 \eta_\text{QCD} \left(
            1-\frac{m_{\pi^+}^2}{m_{K^+}^2}
        \right)^2 \nonumber \\
        & \hspace{5cm} \times \sqrt{1 - \frac{4 m_P^2}{q^2}}
        \sqrt{\lambda\left(1,m_{\pi^+}^2/m_{K^+}^2,q^2/m_{K^+}^2\right)}~,
\end{align}
In the $K^+ \to \pi^+ PP$ decay width, $q^2$ denotes the $PP$ invariant mass, which lies in the range $4 m_P^2 < q^2 < (m_{K^+} - m_{\pi^+})^2$.

The interactions of $S$ and $P$ with quarks that we have introduced preserve a $Z_2$ symmetry under which $S$ and $P$ are odd, while all SM particles are even. We assume that the $Z_2$ symmetry is also respected by the scalar potential, such that $P$ is an absolutely stable DM candidate. Among the allowed $Z_2$ symmetric terms in the scalar potential, the $SP^3$ interaction
\begin{equation}
    \mathcal L_\text{int} \supset \lambda_{SP^3} S P^3 ~,
\end{equation}
will turn out to be relevant. When kinematically allowed, this interaction leads to the decay $S \to 3P$ with rate
\begin{equation}
\Gamma(S \to 3 P) = \frac{3}{256 \pi^3} \lambda_{SP^3}^2 m_S ~  f(m_P/m_S),
\end{equation}
where $f$ is the three-body phase space integral,
\begin{equation}
    f(y) = 2 \int_{4y^2}^{(1-y)^2} \du x ~ \sqrt{\lambda\left(1,x,y^2\right)\lambda\left(1,y^2/x,y^2/x\right)},
\end{equation}
which is normalized to 1 in the limit $y \to 0$. The $S \to 3 P$ rate will modify the lifetime of $S$ and can therefore have a crucial impact on possible constraints from beam-dump experiments.

\subsection{Events at the KOTO experiment} \label{sec:KOTO}
The model introduced in the previous section will lead to $K_L \to \pi^0 PP$ events at the KOTO experiment. We now identify the regions of parameter space in which this decay can mimic the KOTO signal.

\begin{figure}[tb]
\centering
\includegraphics[width=0.48\textwidth]{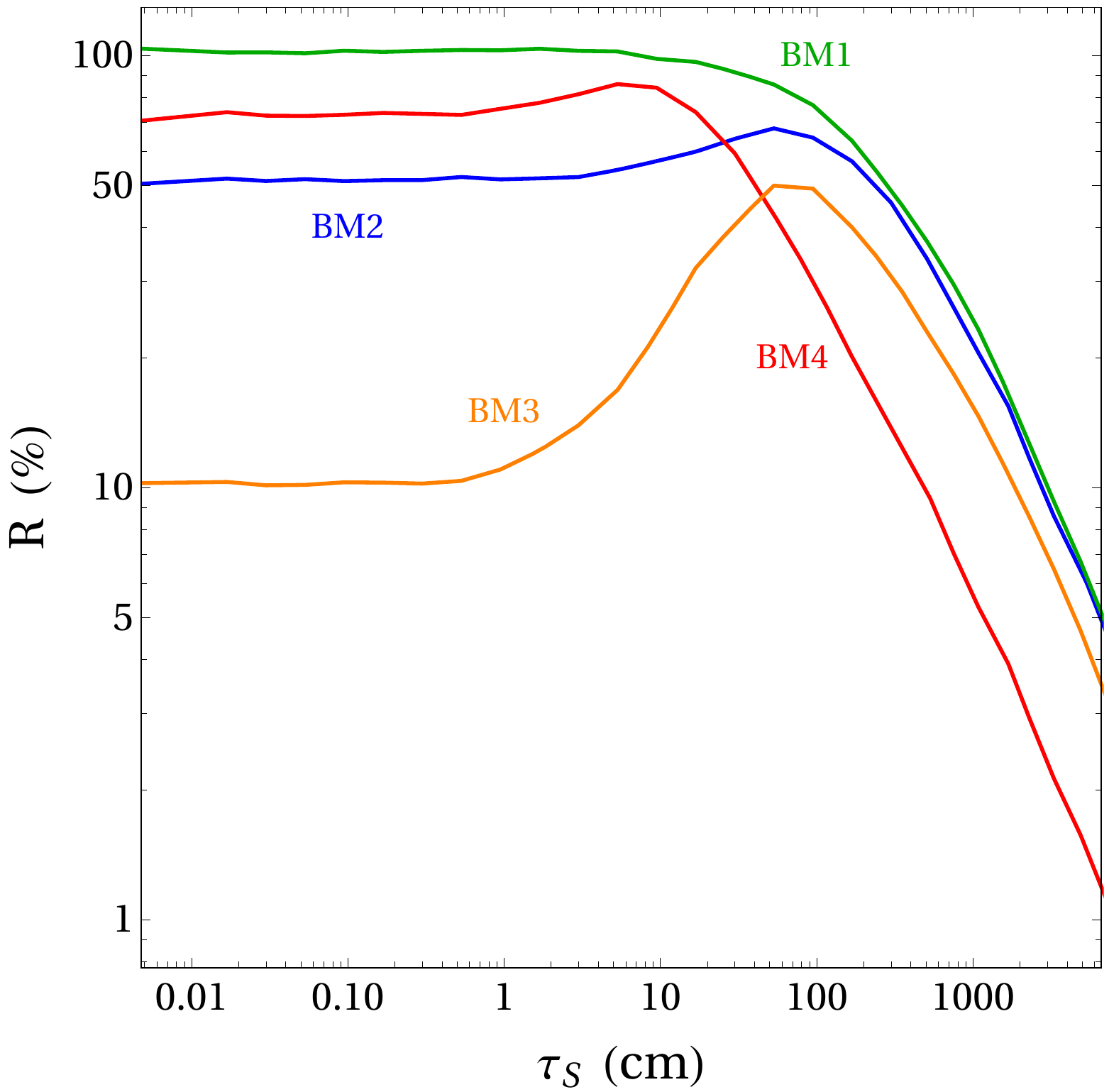} 
\caption{The acceptance ratio $R$ of the $K_L \to SP \to \pi^0 PP$ signal over the SM $K_L \to \pi^0 \nu\bar\nu$ signal at KOTO as a function of the $S$ lifetime $\tau_S$ for the four benchmark scenarios.}
\label{fig:R}
\end{figure}

The number of events that can be expected to be detected at KOTO can be written as
\begin{equation}
    N = \frac{\BR(K_L \to SP) \times \BR(S \to \pi^0 P)}
             {\BR(K_L \to \pi^0 \nu\bar\nu)_\text{SM}}
        \times R \times N_\text{SM},
\end{equation}
where $\BR(K_L \to \pi^0 \nu\bar\nu)_\text{SM} = (3.4 \pm 0.6) \times 10^{-11} $ is the SM prediction for the $K_L \to \pi^0 \nu\bar\nu$ branching ratio~\cite{Brod:2010hi,Buras:2015qea}, $N_\text{SM} = 0.05 \pm 0.01$ is the expected number of SM signal events at KOTO~\cite{KOTO}, and
\begin{equation} \label{eq:R}
 R = \frac{A(K_L \to SP \to \pi^0 PP) }{A(K_L \to \pi^0 \nu\bar\nu) }
\end{equation}
is the ratio of acceptances of the considered model signal and the SM signal at the KOTO detector. As has been pointed out before~\cite{Kitahara:2019lws,Gori:2020xvq,Hostert:2020gou}, an exotic contribution to the KOTO signal (in our case $K_L \to SP \to \pi^0 PP$) can have a considerably different acceptance. We determine the acceptance ratio $R$ using a Monte Carlo simulation. Details are provided in \cref{app:MonteCarlo}. The result is given in \cref{fig:R}, which shows $R$ as a function of the $S$ lifetime for our four benchmark points (\cref{eq:benchmarks}). For prompt decays $\tau_S \to 0$, we find $\{ R_\text{BM1}, R_\text{BM2}, R_\text{BM3}, R_\text{BM4} \} \simeq \{ 102\%, 51\%, 10\%, 73\% \}$. Once the lifetime of $S$ becomes comparable to the size of the KOTO detector, i.e., $\tau_S \gtrsim \SI{1}{\meter}$, $R$ starts to decrease as more and more $S$ leave the detector before decaying.

Note the particularly small acceptance ratio for a promptly decaying $S$ in BM3, $R_\text{BM3} \simeq 10\%$. In fact, the masses of $S$ and $P$ in BM3 are such that the pion's transverse momentum is typically below the experimental cut of $\SI{130}{\MeV}$. However, if $S$ decays not promptly but with a considerable displacement in the detector, the pion momentum is often mis-reconstructed, appears much larger than it is, and therefore more easily passes the cut. This explains the maximum of $R$ at a lifetime of around $\SI{1}{\meter}$. This effect can also be observed, albeit to a much smaller extent, in BM2 and BM4. In BM1, the pion momentum is sufficiently large to pass the cut, independent of the $S$ lifetime.

In our setup, the lifetime of $S$ is determined by the $S \to \pi^0 P$ and $S \to 3P$ decays. In the four benchmark cases for the scalar masses defined above, we find
\begin{multline}
    \Big\{
        \Gamma(S \to \pi^0 P)_\text{BM1},
        \Gamma(S \to \pi^0 P)_\text{BM2},
        \Gamma(S \to \pi^0 P)_\text{BM3},
        \Gamma(S \to \pi^0 P)_\text{BM4}
    \Big\} \simeq
    \\
    \left\{
        \frac{1}{\SI{3.3}{\centi\meter}},
        \frac{1}{\SI{3.4}{\centi\meter}},
        \frac{1}{\SI{4.4}{\centi\meter}},
        \frac{1}{\SI{2.7}{\centi\meter}}
    \right\} \times
    \left(\frac{10^6~\text{GeV}}{\Ldd}\right)^2\left(
        \frac{\alpha_s(10^4~\text{GeV})}{\alpha_s(M)}
    \right)^{8/7}, 
\end{multline}
\begin{multline}
    \Big\{
        \Gamma(S \to 3P)_\text{BM1},
        \Gamma(S \to 3P)_\text{BM2},
        \Gamma(S \to 3P)_\text{BM4}
    \Big\} \simeq
    \\
    \left\{
        \frac{1}{\SI{2.0}{\centi\meter}},
        \frac{1}{\SI{49}{\centi\meter}},
        \frac{1}{\SI{4.3}{\centi\meter}}
    \right\} \times
    \left(\frac{\lambda_{SP^3}}{10^{-5}}\right)^2,
\end{multline}
where in the $S \to \pi^0 P$ decay width we have defined $\Ldd = \LNP/\im(\gqp{dd}{SP})$. Note that $S \to 3P$ is not kinematically allowed in benchmark BM3.
The $S \to \pi^0 P$ branching ratio is given by $\BR(S \to \pi^0 P) = \Gamma(S \to \pi^0 P)/[\Gamma(S \to \pi^0 P)+ \Gamma(S \to 3P)]$.

Finally, we find the following $K_L \to SP$ branching ratios 
\begin{multline}
    \Big\{
        \BR(K_L \to SP)_\text{BM1},
        \BR(K_L \to SP)_\text{BM2},
        \BR(K_L \to SP)_\text{BM3},
        \BR(K_L \to SP)_\text{BM4}
    \Big\} \simeq
    \\
    \Big\{
        1.7,\;
        1.8,\;
        2.3,\;
        4.0
    \Big\} \times 10^{-9} \times \left(
        \frac{10^{12}~\text{GeV}}{\Lsd}
    \right)^2 \left(
        \frac{\alpha_s(\SI{e4}{\giga\electronvolt})}{\alpha_s(M)}
    \right)^{8/7},
\end{multline}
where we have defined $\Lsd = \LNP/\re(\gqp{sd}{SP})$.

\begin{figure}[tb]
    \centering
    \includegraphics[width=0.48\textwidth]{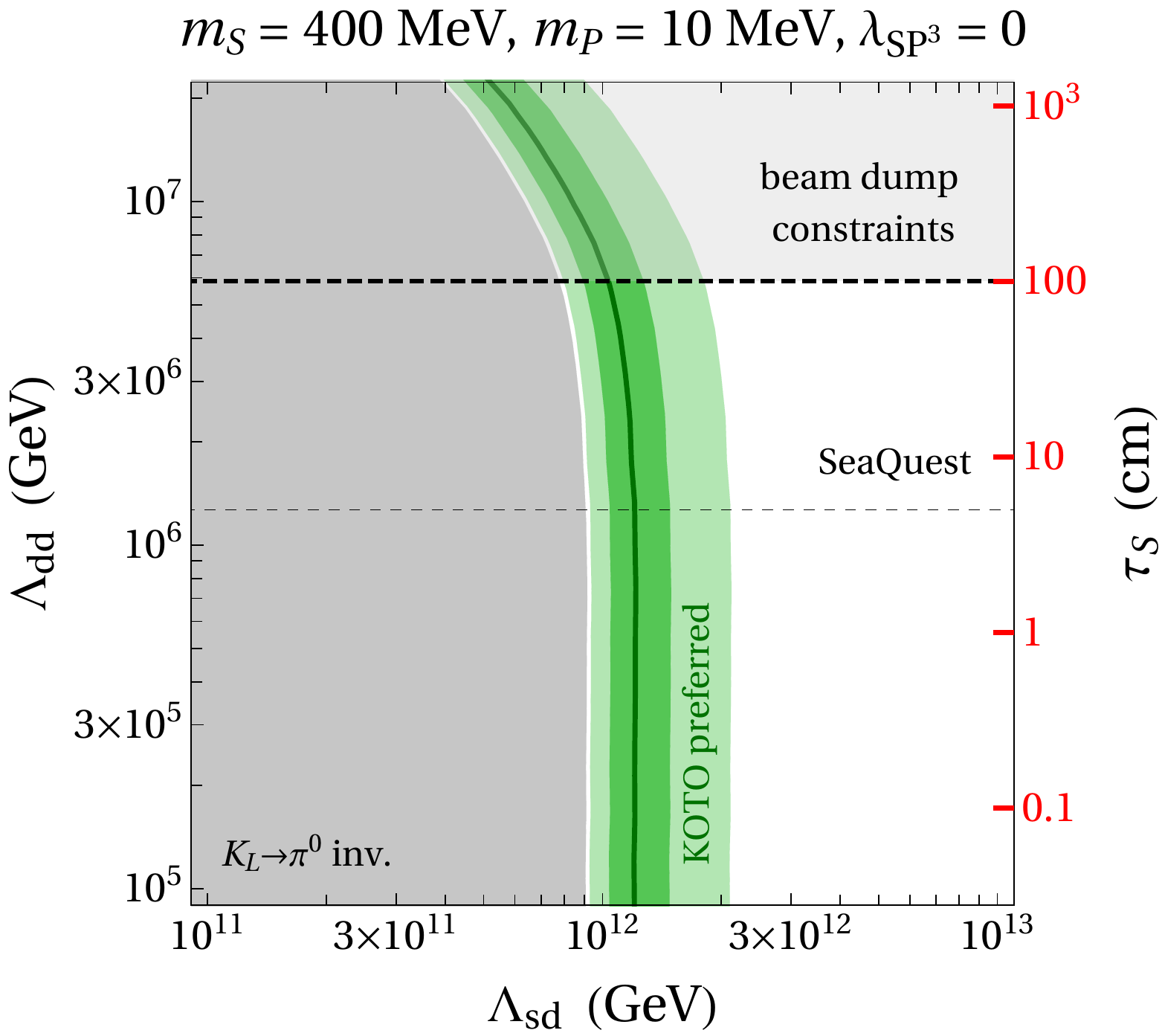} ~~~
    \includegraphics[width=0.48\textwidth]{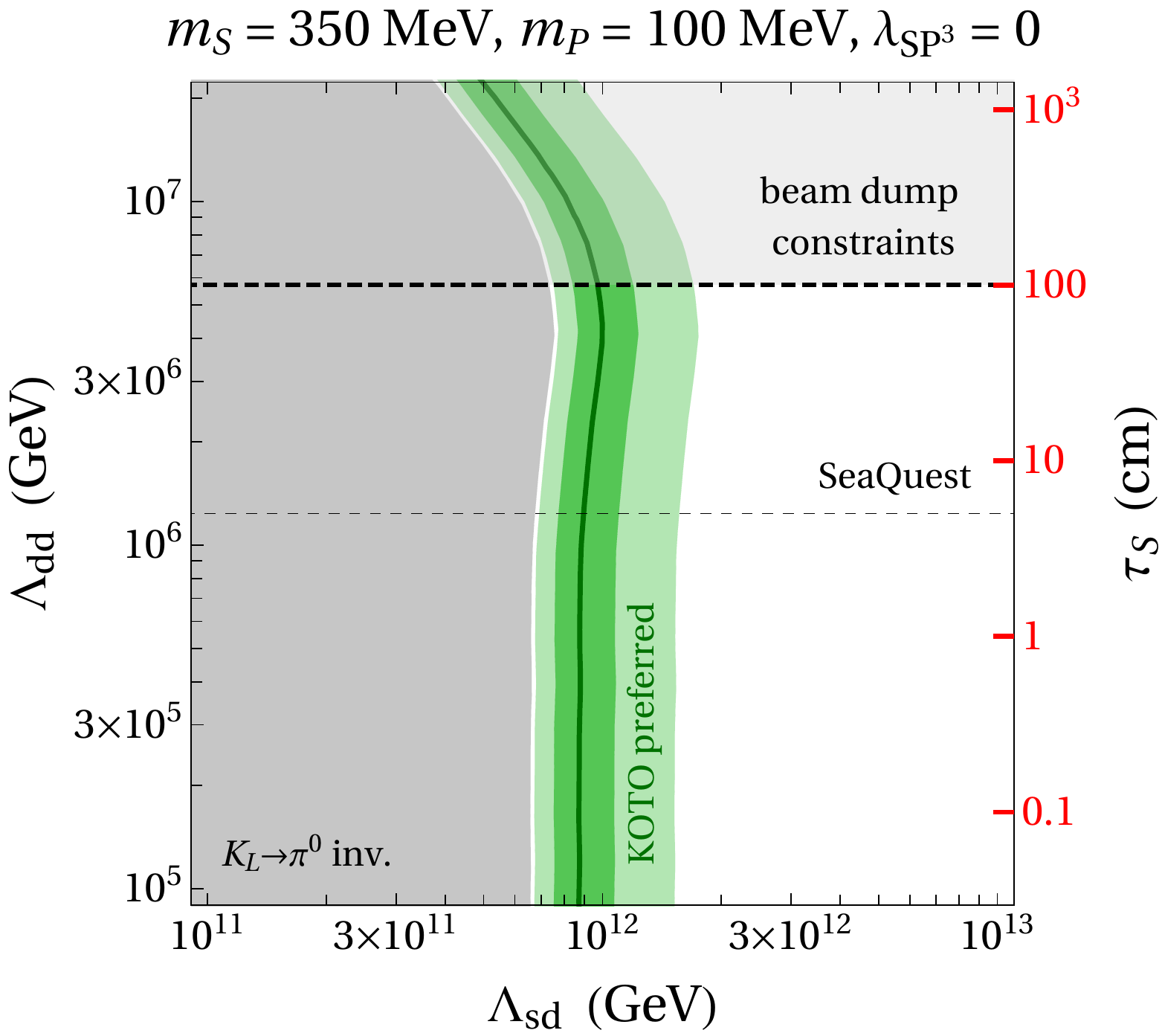} \\[12pt]
    \includegraphics[width=0.48\textwidth]{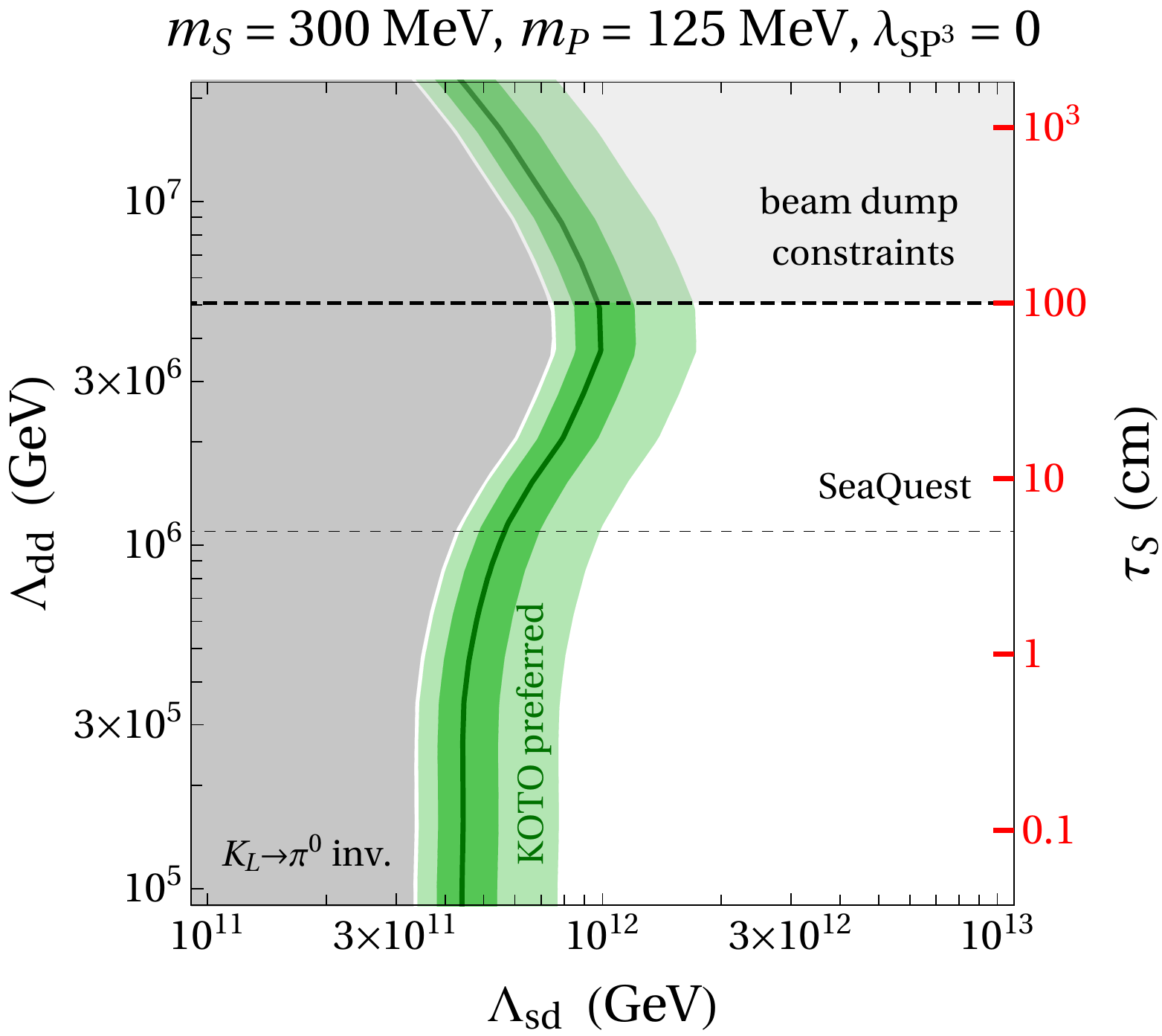} ~~~
    \includegraphics[width=0.48\textwidth]{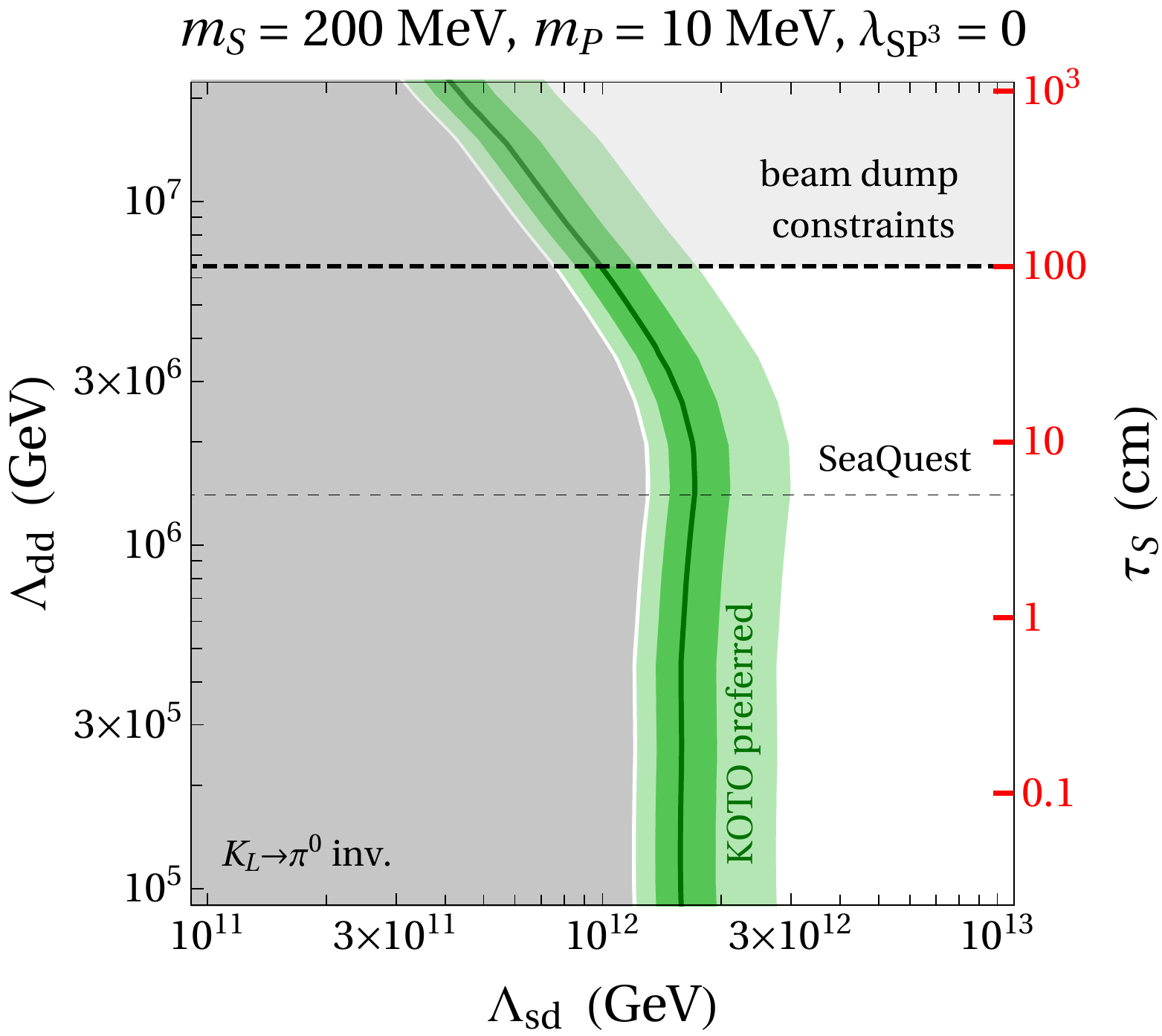} 
    \caption{Number of expected $K_L \to SP \to \pi PP$ events at KOTO in the $\Lsd$--$\Ldd$ plane for four benchmark points of the $S$ and $P$ masses. The $SP^3$ coupling is set to zero. The right vertical axis indicates the $S$ lifetime. One expects 3 events along the solid dark green line, 2--4 events in the dark green region, and 1--5 events in the light green region. In the gray regions labeled ``$K_L \to \pi^0~\text{inv.}$'', the number of predicted events exceeds the limit from KOTO. The dashed lines show constraints from existing beam-dump experiments and the potential reach of the SeaQuest upgrade.}
    \label{fig:Lambda}
\end{figure}

\begin{figure}[tb]
    \centering
    \includegraphics[width=0.48\textwidth]{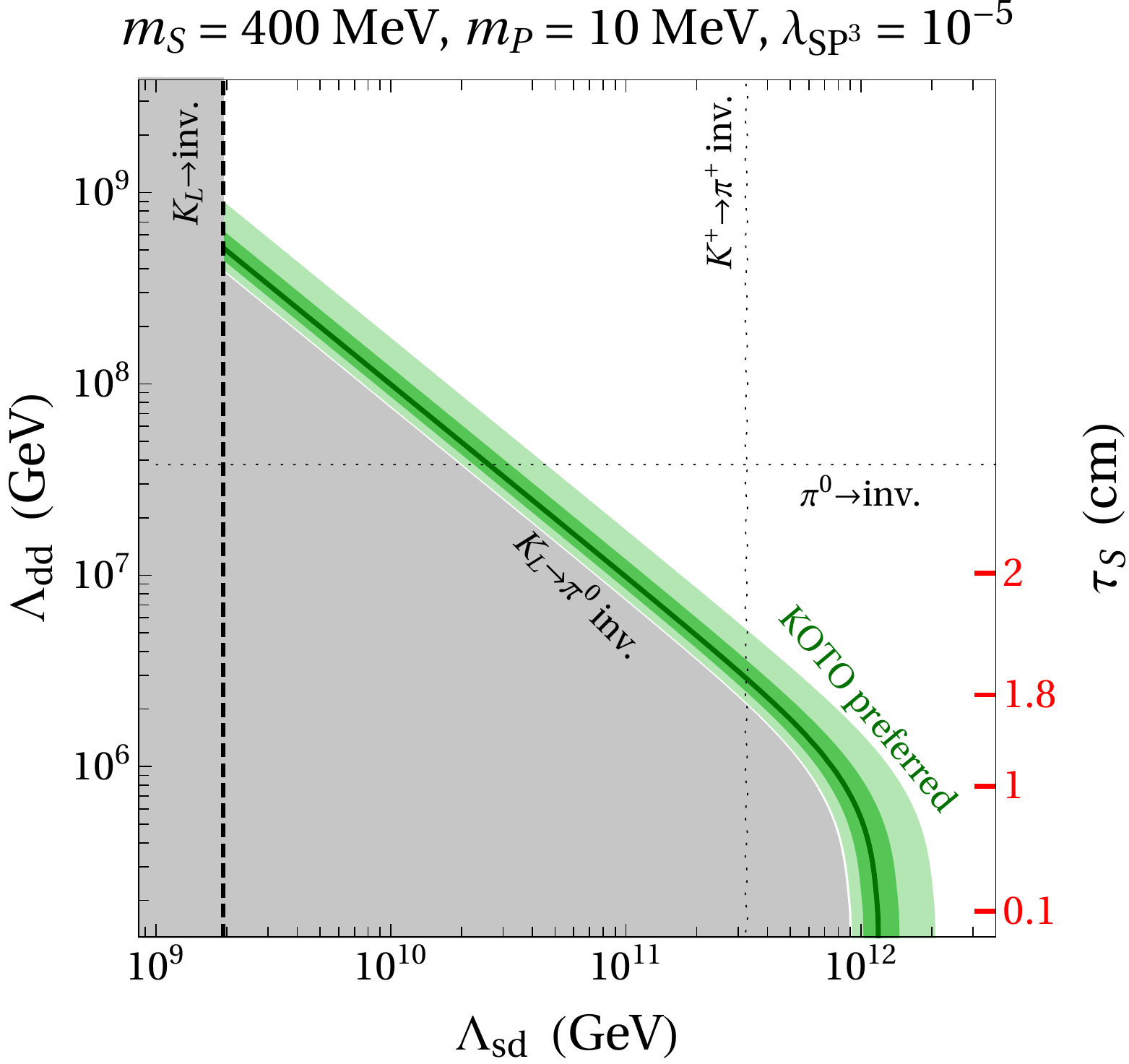} ~~~
    \includegraphics[width=0.48\textwidth]{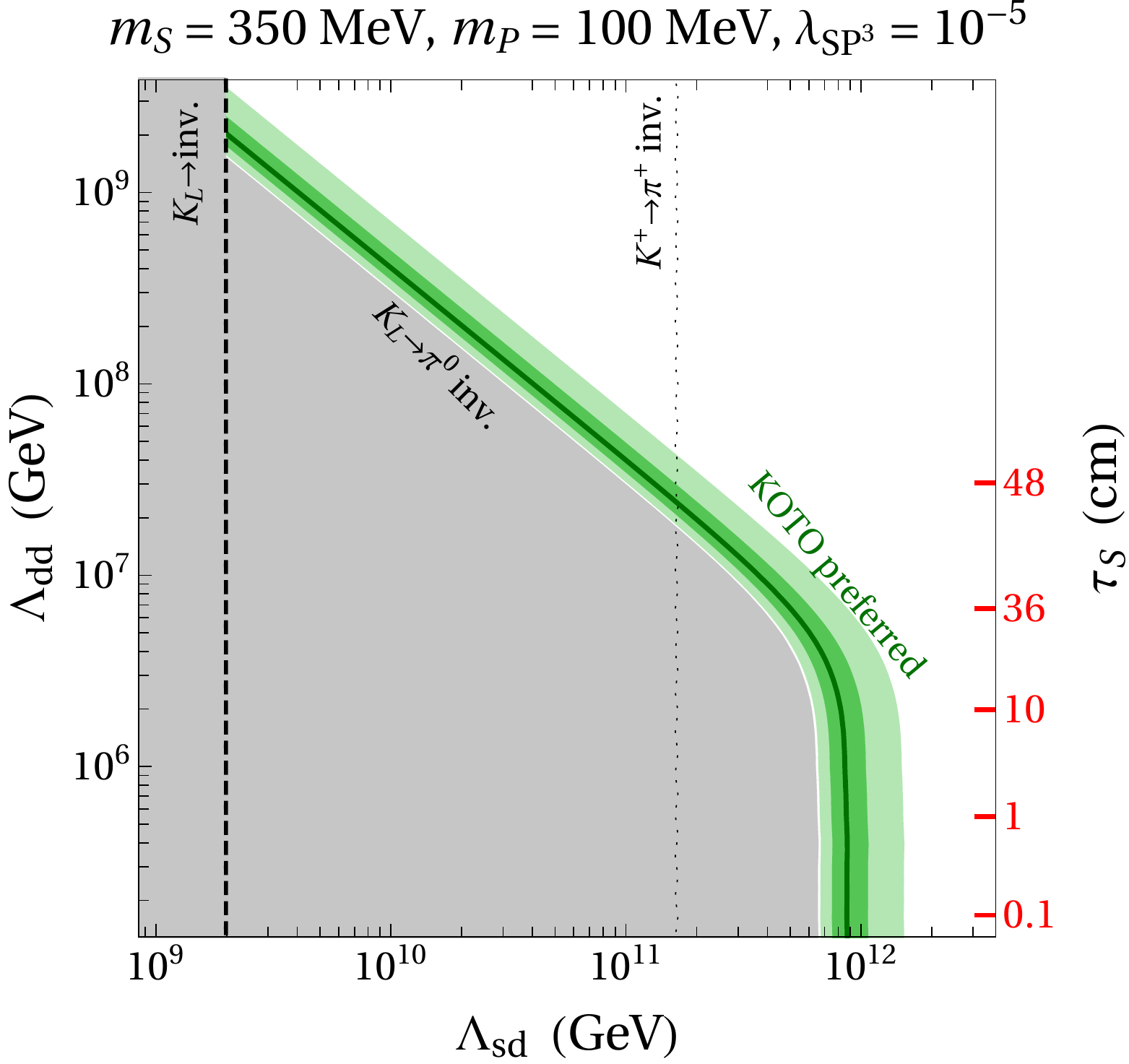} \\[12pt]
    \includegraphics[width=0.48\textwidth]{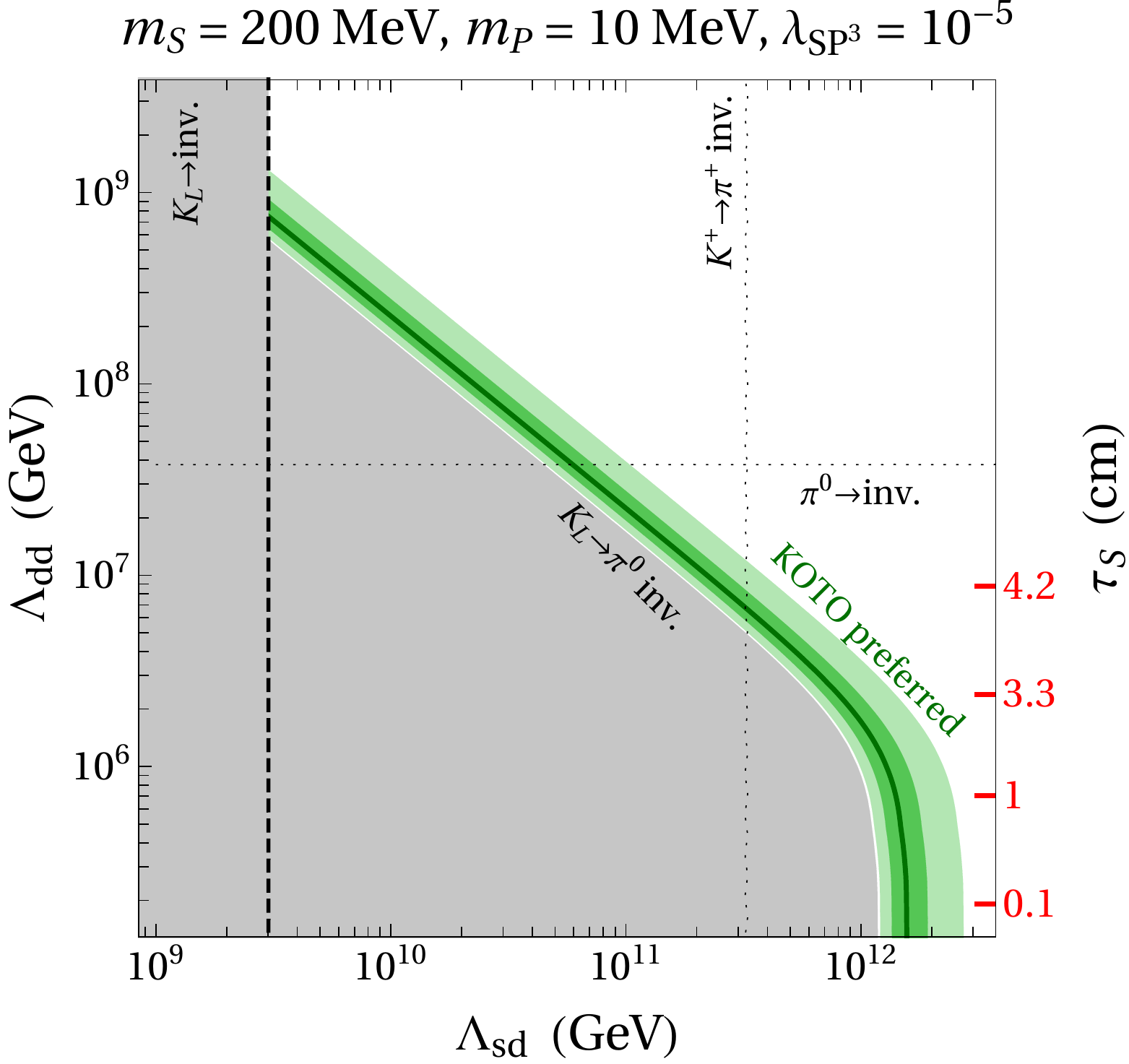} 
    \caption{Number of expected $K_L \to SP \to \pi PP$ events at KOTO in the $\Lsd$--$\Ldd$ plane for three benchmark points of the $S$ and $P$ masses. The $SP^3$ coupling is set to $\lambda_{SP^3} = 10^{-5}$. The right vertical axis indicates the $S$ lifetime, which is approximately constant for $\Ldd > \SI{e7}{\giga\electronvolt}$. One expects 3 events along the solid dark green line, 2--4 events in the dark green region, and 1--5 events in the light green region. The gray regions are excluded by the KOTO limit on $K_L \to \pi^0 ~\text{inv.}$ or the bound on the invisible $K_L$ branching ratio. The dotted lines show the generic location of other constraints that depend on additional model parameters. Benchmark BM3 is not shown, as the $S \to 3P$ decay is kinematically forbidden.}
    \label{fig:Lambda_SP3}
\end{figure}

\Cref{fig:Lambda,fig:Lambda_SP3} show the number of expected events in the $\Lsd$--$\Ldd$ plane for our benchmark cases in the absence of the $S\to 3P$ decay (\cref{fig:Lambda}) and in the presence of the $S\to 3P$ decay induced by a coupling $\lambda_{SP^3} = 10^{-5}$ (\cref{fig:Lambda_SP3}). Along the solid green lines one expects three events, in the dark green regions one expects two to four events, and in the light green regions one expects one to five events. In the gray regions labeled ``$K_L \to \pi^0~\text{inv.}$,'', the number of predicted events exceeds the limit from KOTO (see \cref{eq:KOTOlimit}).
The right vertical axis shows the lifetime of $S$ corresponding to $\Ldd$. In \cref{fig:Lambda_SP3}, the lifetime is approximately constant for $\Ldd > \SI{e7}{\giga\electronvolt}$, as in this region of parameter space, the lifetime is set by the $S\to 3P$ decay width.


For $S$ lifetimes of $\tau_S \gtrsim \SI{1}{\meter}$, existing beam-dump constraints apply (see \cref{sec:constraints}) as indicated in \cref{fig:Lambda} by the dashed contours. A proposed upgrade of the SeaQuest experiment might probe $S$ lifetimes as low as $\tau_S \gtrsim \SI{5}{\centi\meter}$. In the scenarios shown in \cref{fig:Lambda_SP3} with $\lambda_{SP^3} = 10^{-5}$, the $S$ lifetimes are short enough throughout the parameter space that existing beam-dump constraints are avoided.

In \cref{fig:Lambda_SP3} we also show additional constraints from other meson decays.  The known $K_L$ branching fractions add up to a value compatible with $1$ with very high precision. Any additional $K_L$ branching ratio, in particular $K_L \to SP$, is thus bounded above as $\BR(K_L \to SP) < 6.3 \times 10^{-4}$~\cite{Gninenko:2014sxa}. In \cref{fig:Lambda_SP3}, the gray regions left of the dashed vertical lines denoted ``$K_L \to~\text{inv.}$'' are excluded by this constraint. Note that this gives an absolute lower bound $\Lsd \gtrsim \text{few} \times \SI{e9}{\giga\electronvolt}$.

The other meson decay constraints shown in \cref{fig:Lambda_SP3} are less robust as they depend on couplings that are in principle unrelated. If we assume that the coupling $\gq{sd}{P^2}$ [corresponding to $(\bar s d) P^2$] is of the same order as the coupling $\gqp{sd}{SP}$ [corresponding to $(\bar s i \gamma_5 d)SP$], we find relevant constraints from the searches for $K^+ \to \pi^+ \nu\bar\nu$. To evaluate the constraints, we compare the predicted $K^+ \to \pi^+ PP$ branching ratio with the bound from NA62 given in \cref{eq:NA62}. We correct for the different signal acceptances of $K^+ \to \pi^+ PP$ compared to $K^+ \to \pi^+ \nu\bar\nu$ that arise due to kinematical cuts on the missing mass and the charged pion momentum. For the three $P$ masses relevant to our benchmarks, we find the bounds $\BR(K^+ \to \pi^+ PP) < 2.7 \times 10^{-10}$ for $m_P = \SI{10}{\mega\electronvolt}$, $\BR(K^+ \to \pi^+ PP) < 3.5 \times 10^{-10}$ for $m_P = \SI{100}{\mega\electronvolt}$, and $\BR(K^+ \to \pi^+ PP) < 2.4 \times 10^{-9}$ for $m_P = \SI{125}{\mega\electronvolt}$.
Setting $\LNP/|\gq{sd}{P^2}| = \LNP/\re(\gqp{sd}{SP}) = \Lsd$, we find that in \cref{fig:Lambda_SP3}, the regions left of the dotted vertical lines are excluded.

If we assume that the coupling $\gqp{dd}{P^2}$ (corresponding to $(\bar d i\gamma_5 d)P^2$) is of the same order as the coupling $\gqp{dd}{SP}$ (corresponding to $(\bar d i \gamma_5 d)SP$), we find relevant constraints from the invisible branching fraction of the neutral pion, $\BR(\pi^0 \to \text{inv.}) < 4.4\times 10^{-9}$~\cite{NA62talk}. Setting $\LNP/\re(\gqp{dd}{P^2}) = \LNP/\im(\gqp{dd}{SP}) = \Ldd$ in the benchmarks BM1 and BM4, the regions below the dotted horizontal lines are excluded.
For benchmarks BM2 and BM3, the $P$ mass is too large for the $\pi^0 \to PP$ decay, so the couplings are therefore completely unconstrained by $\BR(\pi^0 \to \text{inv.})$.

\subsection{Simplified UV models} \label{sec:UV}

The higher-dimensional interactions in \cref{eq:Leff} that lead to the exotic meson decays can be UV completed by simplified models in various ways. In this section, we discuss briefly two possibilities: (1) vectorlike quarks and (2) an inert Higgs doublet.

\subsubsection{Vectorlike quark model}
We introduce two sets of heavy vectorlike quarks $D$ and $Q$ which have quantum numbers of the right-handed down quark singlets and the left-handed quark doublets, respectively, i.e., $D = ({\mathbf 3},{\mathbf 1})_{-\frac{1}{3}}$ and $Q = ({\mathbf 3},{\mathbf 2})_{\frac{1}{6}}$. These quantum number assignments admit the following terms in the Lagrangian:
\begin{multline}
    \label{eq:VLquarks}
    \mathcal L ~\supset~
        m_Q \bar Q_L Q_R + m_D \bar D_L D_R + Y_{QD} (\bar Q_L D_R) h +
        Y_{DQ} (\bar D_L Q_R) h^c ~+~\text{H.c.} \\
        + X_{Dd} (\bar D_L d_R) S + X_{Ds} (\bar D_L s_R) S +
        Z_{Qd} (\bar Q_R d_L) iP + Z_{Ds} (\bar Q_R s_L) iP
        ~+~\text{H.c.}
\end{multline}
The first line contains the masses $m_Q$ and $m_D$ for the vectorlike quarks, as well as interactions with the SM Higgs doublet $h$. The masses $m_Q$, $m_D$ and the couplings $Y_{QD}$, $Y_{DQ}$ are, in general, complex parameters. However, not all of their phases are observable. Using the freedom to rephase the vectorlike quark fields, we will choose real $m_Q$, $m_D$ and $Y_{QD}$ without loss of generality.
The second line in \cref{eq:VLquarks} contains couplings of the SM down and strange quarks with $S$ and the vectorlike quark $D$ as well as with $P$ and the vectorlike quark $Q$.
The couplings $X_{Dd}$, $X_{Ds}$, $Z_{Qd}$, and $Z_{Qs}$ contain physical phases.

Note that the above Lagrangian is invariant under a $Z_2$ symmetry under which all SM particles are even, while the vectorlike quarks as well as $S$ and $P$ are odd. Thus $P$ remains an absolutely stable DM candidate.
In addition to the couplings shown, the model could also contain $Z_2$ invariant couplings involving $S$ and $Q$ or $P$ and $D$. However, such couplings are not required to generate the desired low-energy interactions, and we will neglect them in the following.

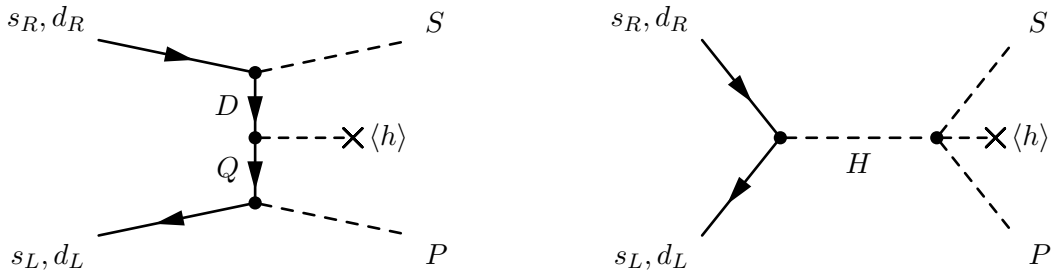
\begin{figure}
    \vspace{1cm}
    \begin{fmffile}{vectorlike-quarks}
        \begin{fmfgraph*}(40,20)
            \fmfpen{thin}
            \fmfleft{i2,i1}
            \fmfright{o3,o2,o1}
            \fmf{fermion}{i1,v1}
            \fmf{fermion,label=$D$}{v1,v2}
            \fmf{fermion,label=$Q$}{v2,v3}
            \fmf{fermion}{v3,i2}
            \fmf{dashes}{v1,o1}
            \fmf{dashes}{v3,o3}
            \fmffreeze
            \fmf{dashes}{v2,vev}
            \fmf{phantom}{vev,o2}
            \fmfdot{v1,v2,v3}
            \fmflabel{$s_R,d_R$}{i1}
            \fmflabel{$s_L,d_L$}{i2}
            \fmflabel{$S$}{o1}
            \fmflabel{$P$}{o3}
            \fmflabel{$\langle h\rangle$}{vev}
            \fmfpen{thick}
            \fmfv{d.sh=cross,d.si=10}{vev}
            \fmfpen{thin}
        \end{fmfgraph*}
    \end{fmffile} \qquad\qquad\qquad
     \begin{fmffile}{inert-higgs}
        \begin{fmfgraph*}(40,20)
            \fmfleft{i2,i1}
            \fmfright{o3,o2,o1}
            \fmf{fermion}{i1,v1}
            \fmf{fermion}{v1,i2}
            \fmf{dashes,label=$H$}{v1,v2}
            \fmf{dashes}{v2,o1}
            \fmf{dashes}{v2,o3}
            \fmffreeze
            \fmf{dashes}{v2,vev}
            \fmf{phantom}{vev,o2}
            \fmfdot{v1,v2}
            \fmflabel{$s_R,d_R$}{i1}
            \fmflabel{$s_L,d_L$}{i2}
            \fmflabel{$S$}{o1}
            \fmflabel{$P$}{o3}
            \fmflabel{$\langle h\rangle$}{vev}
            \fmfpen{thick}
            \fmfv{d.sh=cross,d.si=10}{vev}
            \fmfpen{thin}
        \end{fmfgraph*}
    \end{fmffile}
    \caption{Feynman diagrams that show the matching of the vectorlike quark model (left) and the inert Higgs model (right) onto the effective $SP q q^\prime$ interactions in \cref{eq:Leff}. }
    \label{fig:diagrams}
\end{figure}

Integrating out the vectorlike quarks at tree level (see \cref{fig:diagrams}, left diagram), and matching onto the effective Lagrangian of \cref{eq:Leff}, we find
\begin{align}
     \frac{\gq{dd}{SP}}{\LNP} &=
        \frac{-i Y_{QD} v}{\sqrt{2} m_Q m_D} \im(X_{Dd} Z_{Qd}^*)~, &
        \frac{\gqp{dd}{SP}}{\LNP} &=
            \frac{iY_{QD} v}{\sqrt{2} m_Q m_D} \re(X_{Dd} Z_{Qd}^*)~, \\
    \nonumber
     \frac{\gq{ss}{SP}}{\LNP} &=
        \frac{-iY_{QD} v}{\sqrt{2} m_Q m_D} \im(X_{Ds} Z_{Qs}^*)~, &
        \frac{\gqp{ss}{SP}}{\LNP} &=
            \frac{iY_{QD} v}{\sqrt{2} m_Q m_D} \re(X_{Ds} Z_{Qs}^*)~, \\
    \nonumber
    \frac{\gq{sd}{SP}}{\LNP} &=
        \frac{Y_{QD} v}{\sqrt{2} m_Q m_D} \frac{1}{2} (
            Z_{Qs} X_{Dd}^* - X_{Ds} Z_{Qd}^*
        )~, &
        \frac{\gqp{sd}{SP}}{\LNP} &=
            \frac{Y_{QD} v}{\sqrt{2} m_Q m_D} \frac{i}{2} (
                Z_{Qs} X_{Dd}^* + X_{Ds} Z_{Qd}^*
            )~.
\end{align}
As required by $SU(2)_L$ invariance, the effective interactions $\gq{ij}{SP}/\LNP$ and $\gqp{ij}{SP}/\LNP$ are proportional to the SM Higgs vacuum expectation value $v \simeq\SI{246}{\giga\electronvolt}$. If all couplings $X_{ij}$, $Y_{ij}$, $Z_{ij}$ are of $\mathcal O(1)$, we can expect vectorlike quark masses $m_{Q,D} \sim \sqrt{\LNP v} \sim \SI{e6}{\giga\electronvolt}$. The couplings above are not all independent but obey the relation
\begin{equation}
    |\gqp{sd}{SP}|^2 - |\gq{sd}{SP}|^2 + 2i \re(\gq{sd}{SP} \gqpc{sd}{SP}) =
        \gqp{dd}{SP} \gqpc{ss}{SP} - \gq{dd}{SP} \gqc{ss}{SP} +
        i(\gqp{dd}{SP} \gqc{ss}{SP} + \gqpc{ss}{SP} \gq{dd}{SP}) ~.
\end{equation}
One therefore expects that the flavor changing couplings are of the order of the geometric mean of the flavor conserving couplings.

The vectorlike quarks also give one-loop contributions to kaon mixing. We checked explicitly that those contributions scale as $v^2/(m_Q^2 m_D^2)$ and are completely negligible.

\subsubsection{Inert Higgs doublet model}
In a second scenario, we introduce an inert Higgs doublet $H$ with mass $m_H$, which couples to down and strange quarks, the SM Higgs, and the scalars $S$ and $P$ through the following interactions:
\begin{multline}
    \mathcal L ~\supset~
        m_H^2 H^\dagger H + \lambda_{SP} (H^\dagger h + h^\dagger H) SP \\
        + Y_{dd} (\bar d_L d_R) H + Y_{ds} (\bar d_L s_R) H
        + Y_{sd} (\bar s_L d_R) H ~+~\text{H.c.}
\end{multline}
As in the vectorlike quark scenario, this inert Higgs Lagrangian is invariant under a $Z_2$ symmetry: $S$ and $P$ are odd, while all other particles are even. 
Additional $Z_2$ symmetric quartic couplings of the inert Higgs involving e.g. $S^2$ or $P^2$ are also possible but are not required to generate the low energy interactions in \cref{eq:Leff}, and we neglect them in the following.

Integrating out the inert Higgs at tree level (see \cref{fig:diagrams}, right diagram), and matching onto the effective Lagrangian of \cref{eq:Leff}, we find
\begin{align}
    \frac{\gq{dd}{SP}}{\LNP} &=
        \frac{i\lambda_{SP} v}{\sqrt{2} m_H^2} \re(Y_{dd})~, &
        \frac{\gqp{dd}{SP}}{\LNP} &=
            \frac{i\lambda_{SP} v}{\sqrt{2} m_H^2} \im(Y_{dd})~,\\
    \frac{\gq{ds}{SP}}{\LNP} &=
        \frac{\lambda_{SP} v}{\sqrt{2} m_H^2} \frac{i}{2}( Y_{ds} +
            Y_{sd}^*)~, &
            \frac{\gqp{ds}{SP}}{\LNP} &=
                \frac{\lambda_{SP} v}{\sqrt{2} m_H^2}\frac{1}{2} (
                    Y_{ds} - Y_{sd}^*
                )~.
\end{align}
In addition, integrating out the inert Higgs gives four-fermion contact interactions of the type $(\bar d_L s_R)(\bar d_R s_L)$ that modify kaon oscillations. We find the following contributions to the kaon mixing matrix element:
\begin{equation}
    M_{12} = \frac{m_{K^0}^3 f_K^2}{4 m_s^2 m_H^2}
        \eta_\text{QCD} B_4 Y_{sd} Y^*_{ds} ~,
\end{equation}
where $B_4 \simeq 0.78$~\cite{Carrasco:2015pra} (see also~\cite{Jang:2015sla,Garron:2016mva}), and $\eta_\text{QCD}$ is the QCD correction factor given in \cref{eq:QCDrunning}, with $M = m_H$. Modifications to the mixing matrix alter the neutral kaon oscillation frequency $\Delta M_K$ and the observable $\epsilon_K$ that measures $CP$ violation in kaon mixing. The above contribution to $M_{12}$ modifies these two quantities as
\begin{equation}
    \Delta M_K = \Delta M_K^\text{SM} + 2 \re(M_{12}) ~,
    \qquad
    \epsilon_K = \epsilon_K^\text{SM} +
        \frac{\im(M_{12})}{\sqrt{2} \Delta M_K} ~.
\end{equation}
Taking into account the SM predictions $\Delta M_K^\text{SM}$ and $\epsilon_K^\text{SM}$ from~\cite{Brod:2011ty,Brod:2019rzc}, and the corresponding experimental values from~\cite{Tanabashi:2018oca}, we find the bounds
\begin{align}
    \re(Y_{sd}Y_{ds}^*) &< 7.3 \times 10^{-9} \times \left(
        \frac{m_H}{1~\text{TeV}}
    \right)^2 \left(
        \frac{\alpha_s(m_H)}{\alpha_s(1~\text{TeV})}
    \right)^{8/7} ~, \\
    \im(Y_{sd}Y_{ds}^*) &< 4.5 \times 10^{-12} \times \left(
        \frac{m_H}{1~\text{TeV}}
    \right)^2 \left(
        \frac{\alpha_s(m_H)}{\alpha_s(1~\text{TeV})}
    \right)^{8/7} ~.
\end{align}
Assuming $|Y_{ds}| \simeq |Y_{sd}|$ and $\mathcal O(1)$ $CP$-violating phases, the kaon mixing bounds are compatible with $\Lsd \gtrsim 3\times\SI{e9}{\giga\electronvolt}$. Also, note that the bounds are entirely avoided if either of $Y_{sd}$ or $Y_{ds}$ is set to zero.

\section{Astrophysical and terrestrial constraints} \label{sec:constraints}
We now consider extant astrophysical and terrestrial constraints that may apply to our model.

First, anticipating our treatment of $P$ as a DM candidate, we note that direct detection, indirect detection, and self-interaction constraints are not relevant for our model in its minimal configuration [see \cref{eq:Leff}]. If our $P$ is the cosmological DM, but the SM is only coupled to the current $SP$, then direct detection is only sensitive to the inelastic scattering process $P + \mathrm{SM} \to S + \mathrm{SM}$, which is kinematically forbidden unless the DM is boosted. Similarly, indirect detection and self-interaction processes require two vertices, and thus the cross sections are suppressed by $\LNP^4$.

Extensions of our minimal model containing couplings to $P^2$ [see \cref{eq:Leff2}] may be subject to these constraints due to the presence of additional interactions. However, we first treat constraints from supernova cooling and beam-dump experiments, which apply directly to the minimal model.

\subsection{Supernova constraints} \label{sec:supernova}
Supernova cooling provides powerful constraints on new weakly coupled light particles. Evaluating these bounds properly requires a detailed analysis that lies beyond the scope of this work, but we can perform an order-of-magnitude estimate to determine the regions of our parameter space that are likely to be subject to such constraints.

In the case of axions, the cross section for axion production $NN\to NNa$ is constrained by SN1987A to lie in the range \cite{Raffelt:1996wa}
\begin{equation}
    \label{eq:sn-bound}
    3\times 10^{-20}\lesssim \frac{\sigma}{{\rm GeV}^{-2}}\lesssim 10^{-13}.
\end{equation}
Below the lower limit, axions are not produced in sufficient numbers to affect the cooling process. Above the upper limit, the produced axions are trapped within the supernova environment, and are unable to cool the system more effectively than neutrinos. Many details of the calculation for axions should be modified in our case, but we will make a crude estimate of the constraints by requiring our production cross section to lie in the same range. We focus on processes with $P$ in the final state: since $m_S>m_P$, the constraints from production of $S$ are typically weaker.

Since $P$ is stabilized by a $Z_2$ symmetry, it can only be produced in pairs, or in association with $S$. The process $NN\to NNPP$ is mediated at the loop level in the minimal model, involving two insertions of the effective interaction vertex. Since $\TSN \simeq \SI{30}{\mega\electronvolt}$ \cite{Raffelt:1996wa}, we estimate the cross section as
\begin{equation}
    \sigma_{NN\to NNPP} \sim
        \frac{1}{16 \pi^2} \frac{T_{\rm SN}^2}{\Ldd^4}
        \simeq
        \SI{6e-34}{\giga\electronvolt^{-2}}\left(
            \frac{\TSN}{\mathrm{30\ MeV}}
        \right)^2\left(
            \frac{\SI{e7}{\giga\electronvolt}}{\Ldd}
        \right)^4,
\end{equation}
lying below the constrained range of cross sections, even neglecting exponential suppression when $m_P\gtrsim \TSN$. In the case of $SP$ production, $NN\to NNSP$, since $m_S \gg \TSN$, we estimate the cross section as 
\begin{multline}
    \sigma_{NN\to NNSP} \sim
    \frac{1}{4\pi \Ldd^2}e^{-(m_S+m_P)/T_{\rm SN}} \\
    \simeq
    \SI{7e-21}{\giga\electronvolt^{-2}}\exp\left[
        \frac{35}{3}\left(
            1 - \frac{m_S + m_P}{\SI{350}{\mega\electronvolt}}
            \frac{\SI{30}{\mega\electronvolt}}{\TSN}
        \right)
    \right] \left(\frac{\SI{e7}{\giga\electronvolt}}{\Ldd}\right)^2
    ,
\end{multline}
typically below the reach of the constraint of \cref{eq:sn-bound}. For a full accounting of supernova constraints, one should also consider processes involving the decay of an off-shell $S$, such as $NN\to NN\pi PP$ and $NN\to NNPPPP$. The former is suppressed by an additional factor of the effective interaction scale, as with $NN\to NNPP$. The latter, on the other hand, proceeds through the $SP^3$ interaction, which is not necessarily suppressed by any large scales. For smaller values of $m_P$, this contribution may be significant. However, as the coupling in this interaction can be adjusted independently of the others, such constraints are model-dependent.

In summary, parts of our parameter space are expected to be unconstrained by supernova limits, but it is important to note that if $m_P$ is small, or if $\Ldd\lesssim\SI{e6}{\giga\electronvolt}$, the estimated production cross section enters the prohibited range. In particular, if $\Ldd = \SI{e6}{\giga\electronvolt}$, then avoiding the bound requires $m_S + m_P \gtrsim \SI{450}{\mega\electronvolt}$, favoring the larger $P$ masses in \cref{fig:masses}. However, in this naive projection of supernova constraints, our model remains viable in a wide region of the parameter space.

\subsection{Beam dump constraints} \label{sec:beam-dump}
In minimal form, our model of the KOTO excess is potentially subject to constraints from long-lived particle searches: The partial decay width of $S\to \pi^0P$ is bounded from below by the observed KOTO event rate, so in the absence of additional interactions, the $S$ lifetime can be $\mathcal O(\SI{}{\meter})$ or larger. Such lifetimes are probed very effectively by beam-dump experiments with $\mathcal O(\SI{100}{\meter})$ baseline lengths.
In such an experiment, a proton beam is directed at a target, potentially producing a large number of $S$ particles. The $S$ particles travel unimpeded through shielding and earth over a distance $L_B$, reaching an instrumented decay volume with length $L_D$. The $S\to \pi^0P$ events within the decay volume can be typically detected with an $\mathcal O(1)$ efficiency $\mathcal E$. Thus, the strength of the constraints is mainly determined by two factors: (1) how many $S$ particles are produced, and (2) what fraction of these undergo $S\to\pi^0P$ within the decay volume.

First we estimate the number of $S$ particles produced. There are at least two channels to consider: direct production from nucleon-nucleon scattering, and secondary production from kaon and other meson decays. Observe, however, that the fraction of proton-proton collisions that produce an $SP$ pair is of order $(s/\LNP)^2/\alpha_S^2$, which is much smaller than the branching ratios $\BR(K_L\to SP)$ and $\BR(K_S\to SP)$ implied by our interpretation of the KOTO excess. We also checked that the number of $S$ from eta decays $\eta \to SP$ is small compared to those coming from the kaon decays in our scenarios. 

Given $N_p$ protons on target, we expect that of order $N_K\sim 10^{-2} N_p$ kaons are produced \cite{Gori:2020xvq}, and this is sufficient for kaon decays to dominate production. However, of these kaons, most will be stopped or scattered away from the axis of the beam before they decay. The dynamics of kaon energy loss and deflection in materials are complicated, but the nuclear interaction length for relativistic kaons in most materials is $L_{\text{nuc}}\sim\mathcal O(\SI{10}{\centi\meter})$ \cite{Tanabashi:2018oca}, so we will assume that any kaons traveling this far before decaying are sufficiently slowed down or deflected such that only a negligible fraction of the $S$ particles are directed toward the detector. Thus, the number of $S$ particles produced and directed toward the detector is of order
\begin{equation}
    N_S \sim \frac{1}{2}\sum_{X=L,S}10^{-2}N_p\frac{\Gamma(K_X\to SP)}{\Gamma_{K_X}}\left[
        1-\exp\left(
            -\frac{\Gamma_{K_X}L_{\text{nuc}}}{\gamma_{K_X}}
        \right)
    \right],
\end{equation}
where $\gamma$ is the boost factor. Now, accounting for the fraction of $S$ particles which decay in the decay volume, and accounting for the efficiency of the detector, the number of events is given by
\begin{multline}
    N_E \sim \frac{1}{2}\sum_{X=L,S}
    10^{-2}N_p\BR(K_X\to SP)\BR(S\to\pi^0P)\mathcal E\\
    \times
    \underbrace{\left[
        1-\exp\left(
            -\frac{\Gamma_{K_X}L_{\text{nuc}}}{\gamma_{K_X}}
        \right)
    \right]}_{\text{avoid kaon deflection}}
    \;
    \underbrace{\exp\left(
        -\frac{\Gamma_{S}L_B}{\gamma_{S}}
    \right)}_{\text{reach decay volume}}
    \;
    \underbrace{\left[
        1-\exp\left(
            -\frac{\Gamma_{S}L_D}{\gamma_{S}}
        \right)
    \right]}_{\text{decay in decay volume}}.
\end{multline}
In the minimal scenario, with no additional interactions, $\BR(S\to\pi^0P)=1$.

We now estimate the event counts in the CHARM \cite{Bergsma:1985qz} and NuCal \cite{Blumlein:1990ay} experiments. CHARM conducted a search for decays of axionlike particles with $2.4\times10^{18}$ protons incident on a copper target at \SI{400}{\giga\electronvolt}, a baseline length of \SI{480}{\meter}, and a \SI{35}{\meter}-long decay volume. The detector efficiency is approximately $0.5$. No candidate events were observed. NuCal conducted a similar search, with $1.7\times10^{18}$ protons incident on an iron target at \SI{70}{\giga\electronvolt}, a baseline length of \SI{64}{\meter}, and a \SI{23}{\meter}-long decay volume. One candidate event was observed with an expected Standard Model background of 0.3. To estimate the event counts that would be produced by our model, we set $\gamma_{K_X}=\gamma_S=10$ for CHARM and reduce these proportionally for NuCal's lower beam energy.

\begin{figure}\centering
    \includegraphics{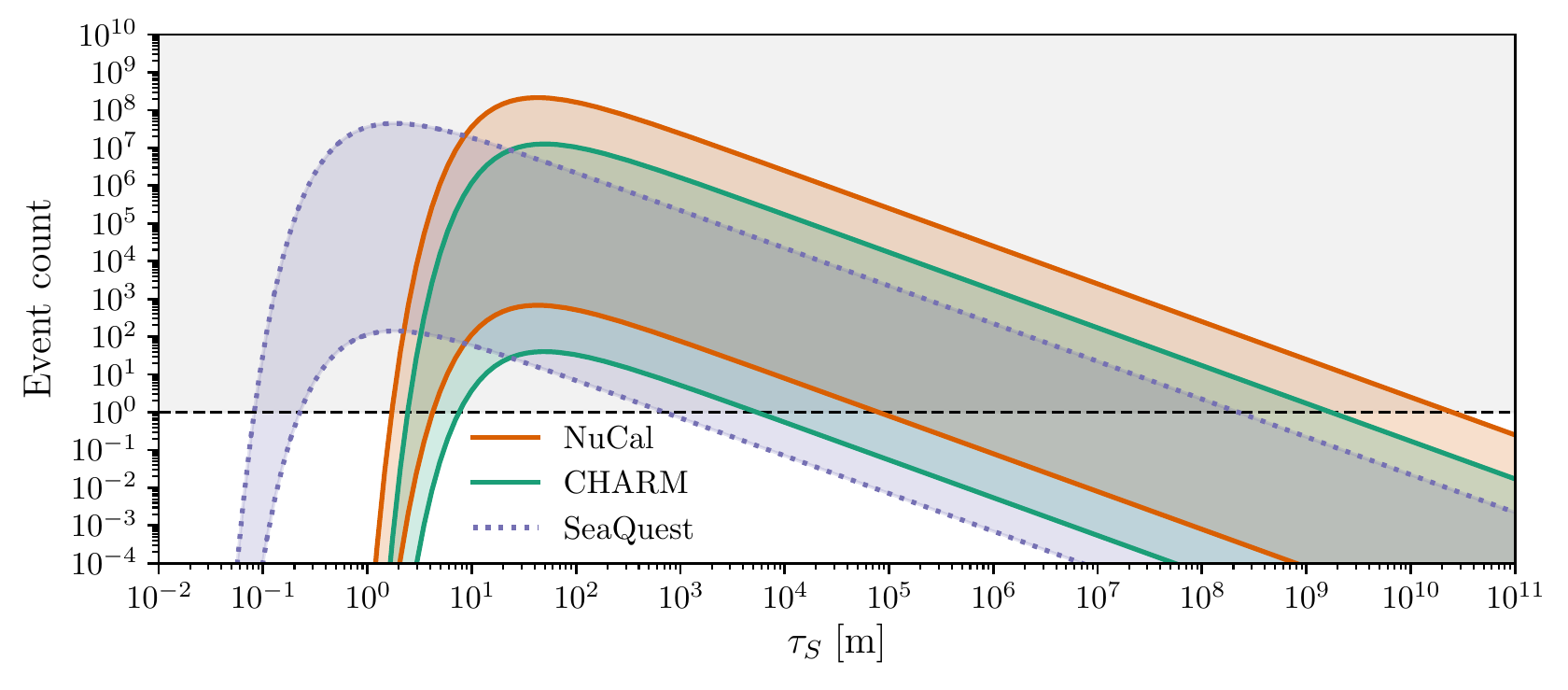}
    \caption{Estimated event counts at CHARM and NuCal and prospective event counts at SeaQuest as a function of the $S$ lifetime. The top curve fixes $\Gamma(K_L\to SP)$ to saturate the experimental bound on the invisible $K_L$ width. The bottom curve fixes $\Gamma(K_L\to SP)$ such that $\BR(K_L\to SP)$ is equal to the ratio inferred from the KOTO excess, i.e., $\Gamma(K_L\to SP)$ is the smallest width for which this model can account for the excess. Both curves assume that $\Gamma(K_L\to SP) = \Gamma(K_S\to SP)$ and that $\BR(S\to\pi^0P)=1$. Under these conditions, $\SI{1}{\meter}\lesssim\tau_S\lesssim\SI{e5}{\meter}$ is ruled out. SeaQuest may eventually probe lifetimes as short as $\tau_S\sim\SI{5}{\centi\meter}$.}
    \label{fig:beam-dump-events}
\end{figure}

Assuming $\BR(S\to\pi^0P)=1$, the resulting event count is shown as a function of the $S$ lifetime in \cref{fig:beam-dump-events}. The minimum expected number of events at long $S$ lifetime is large unless $\tau_S\gtrsim\SI{e5}{\meter}$, and lifetimes as large as \SI{e9}{\meter} may be excluded. This potentially rules out a significant portion of our parameter space, as indicated in \cref{fig:Lambda}. On the other hand, the event rate cuts off sharply for $\tau_S\lesssim\SI{1}{\meter}$, and there is indeed a region of our parameter space where $\tau_S\sim\SI{1}{\centi\meter}$. These constraints can be relaxed if the coupling of the $SP^3$ interaction in our model is nonzero, which can shorten the $S$ lifetime significantly if $m_P$ is small (see \cref{fig:Lambda_SP3}). The presence of this additional interaction greatly extends the parameter space consistent with the null results at CHARM and NuCal.

Looking toward future prospects, most proposed beam-dump experiments are competitive in the same regime of $S$ lifetimes. However, it has been suggested \cite{Berlin:2018pwi} that the SeaQuest experiment \cite{Aidala:2017ofy} may be modified to serve as a short-baseline beam-dump experiment, with the instrumented area starting only $\sim\SI{5}{\meter}$ from the beam target. Such an experiment would have sensitivity to lifetimes as short as \SI{5}{\centi\meter}, and could probe most of the parameter space in which the minimal model can account for the KOTO excess. However, if the $SP^3$ coupling is unconstrained, the $S$ lifetime can be shortened by many orders of magnitude, potentially evading even these experiments.

\subsection{Direct dark matter detection}\label{sec:direct-detection}
Direct detection of $P$ can occur in the extended model via the interactions in \cref{eq:Leff2}. While the interaction terms containing $(\bar qi\gamma_5 q) P^2$ give rise to suppressed velocity-dependent cross sections off of nucleons, the operators $(\bar q q)
P^2$ with $q=d,s$ produce potentially detectable scattering off of nucleons. We define the integrated nucleon form factors
\begin{equation}
    B^N_q\equiv \langle N|\bar q q|N\rangle=\frac{m_N}{m_q}f^N_q,
\end{equation}
where $f^N_q$ are the form factors for nucleon $N$ of quark $q$ \cite{Kelso:2014qja}. The direct detection cross section can be cast as
\begin{equation}
    \sigma=\sum_{q=d,s}\left(
        \frac{2m_N}{m_P+m_N}\frac{\gq{qq}{P^2}}{\LNP}B^N_q
    \right)^2\approx
    \frac{4}{\LNP^2}\left[
            (B_d^N)^2(\gq{dd}{P^2})^2+(B_s^N)^2(\gq{ss}{P^2})^2
    \right].
\end{equation}
Using the central values $B_d^p\approx 6.77$ and $B_s^p\approx 0.50$, it is clear that the dominant effect is scattering off of $d$ quarks if $\gq{ss}{P^2}\simeq \gq{dd}{P^2}$. The scattering cross section off of protons is then
\begin{equation}
    \sigma_p\approx \SI{7e-38}{\centi\meter^2} (\gq{dd}{P^2})^2\left(
        \frac{10^6\  {\rm GeV}}{\LNP}
    \right)^2,
\end{equation}
i.e., close to \SI{0.1}{\pico\barn}. Cross sections of this order are above the expected  neutrino background, and are within reach of future planned experimental sensitivity \cite{Dutta:2019oaj}. We will return to direct detection prospects in \cref{sec:discussion}.

\section{Cosmological production} \label{sec:cosmological-production}
We now turn to the question of cosmological production of the DM candidate $P$: Which scenarios allow $P$ to be produced with the observed DM density?

The standard thermal freeze-out paradigm is not viable in our minimal scenario. Estimating the freeze-out temperature by $n_P\sigma(PP\to\mathrm{SM})\sim H(T)$, we have 
\begin{equation}
    \TFO \sim \frac{4\pi\LNP^2}{\mpl} \sim
        \left(\frac{\LNP}{\SI{e12}{\mega\electronvolt}}\right)^2
        \SI{e3}{\mega\electronvolt},
\end{equation}
where $\LNP$ is the scale of new physics in question---for practical purposes, the lesser of $\Lsd$ and $\Ldd$. For typical values of $\LNP$ consistent with the KOTO excess, $\TFO\gg m_P$, so $P$ freezes out as a hot relic, with relic abundance
\begin{equation}
    \Omega_Ph^2 \sim \frac{m_P}{\SI{}{\kilo\electronvolt}}\left(
        \frac{\left.g_\star\right|_{\TFO}}{100}
    \right) \sim 0.1\left(\frac{m_P}{\SI{80}{\electronvolt}}\right).
\end{equation}
Thus, for the masses and couplings considered in this work, $P$ is generically overproduced in the freeze-out scenario. If the $P$ mass were small enough to be produced with the right relic abundance, then $P$ would be ruled out as a DM candidate because of structure formation constraints on relativistic relics.

Departing from the minimal scenario outlined above opens up the possibility  that an {\em additional} effective interaction with SM species keeps $P$ in thermal equilibrium, and that the $P$ relic abundance is set by thermal decoupling (freeze-out). Since generally thermal decoupling happens at temperatures  $T\sim m_P/25$, in order to avoid possible constraints from big bang nucleosynthesis (BBN), one can assume that the effective interaction only involves SM neutrinos:
\begin{equation}
    {\cal L}\supset \frac{1}{\Lnn}\bar\nu\nu PP.
\end{equation}
For the effective dimension-five operator in the equation above, we find that the zero-velocity thermally averaged product of the pair-annihilation cross section and relative velocity is
\begin{equation}
    \lim_{v\to 0} \langle\sigma v\rangle=\frac{1}{4\pi}\frac{1}{\Lnn^2}.
\end{equation}
A standard treatment of the relic abundance for the pair-annihilation cross section above indicates that $P$ would be produced in the right amount if $\Lnn\simeq 7$ TeV. This is several orders of magnitude above current limits for DM interactions with SM neutrinos, independent of flavor \cite{Blennow:2019fhy}. Thus, if $P$ were in equilibrium at high temperatures, an effective interaction with SM neutrinos---which, incidentally, can be quite naturally embedded in the UV completions described above---could suppress the $P$ abundance to an acceptable relic density in agreement with observations.

In the absence of the additional neutrino portal described in the paragraph above, the only alternative is production via freeze-in \cite{Hall:2009bx}. Here the dark species is produced out of equilibrium by some Standard Model species, and the abundance increases until cosmological expansion halts production. It is thus possible to avoid overproduction of DM with extremely small couplings.
Note that while other mechanisms might allow for additional production of $P$, the freeze-in contribution is unavoidable in the range of temperatures where our effective theory is valid.

Typically, freeze-in is applied to a UV-complete theory, where the DM production rate can be computed starting at very high temperatures. In the context of a renormalizable model, it can be shown that DM is produced primarily at lower temperatures, so the details of the UV physics are unimportant. Thus, freeze-in can be used to consistently calculate the nonthermal relic abundance, even though a formal dependence on initial conditions remains. Note that this is in contrast to the freeze-out paradigm, where equilibrium with the Standard Model bath erases any nontrivial initial conditions in the dark sector.

However, in our scenario, the DM is produced through nonrenormalizable interactions, and the standard freeze-in mechanism cannot be directly applied: Our effective theory cannot be applied at scales above some $\mathcal O(\LNP)$ cutoff. At first, this does not seem to be a problem: In standard freeze-in, production is IR dominated, and we can apply our effective theory in this regime. But for higher-dimension operators, production is no longer IR dominated, and it is no longer possible to self-consistently estimate the relic abundance unless an initial condition is fixed at a temperature where the effective theory is valid.

Naively, one can place a lower bound on the relic abundance by fixing the DM abundance to zero at $T\sim\LNP$ and computing the amount of DM produced at lower temperatures, where the effective theory is valid. However, as we shall see in the following section, this still leads to overproduction of $P$. Thus, in our model, it would seem that DM is overproduced in the freeze-in scenario, even with the most favorable initial conditions.

There is, however, a significant loophole in this argument: Setting the DM abundance to zero at $T\sim\LNP$ is, in fact, not the most favorable initial condition. If reheating occurs at a temperature $\TRH\ll\LNP$, then the DM abundance should be set to zero at this lower temperature, allowing for a much lower relic abundance. There is nothing particularly unnatural about this scenario: In general, freeze-in production of DM depends on the reheating temperature. This dependence is weak if the reheating scale happens to be much higher than any scale in the theory, but while such an arrangement is convenient, there is no direct evidence that this is the case. Moreover, if $\TRH\ll\LNP$, then our effective theory can be used to self-consistently compute the DM relic abundance independent of any UV completion. This paradigm is known as UV freeze-in \cite{Elahi:2014fsa}.

\subsection{Computing the yield}
First, we briefly review the computation of the DM relic abundance in the standard freeze-in paradigm. The basic technology of UV freeze-in is identical to that of standard freeze-in, but the initial condition is fixed at the reheating temperature $\TRH$, which becomes an important free parameter of the theory. In certain scenarios, the DM yield is quite sensitive to temperatures near $\TRH$, and decreasing $\TRH$ can significantly reduce the relic abundance.

The starting point is the Boltzmann equation,
\begin{equation}\label{eq:boltzmann}
    \dot n_\chi + 3Hn_\chi = 
        \sum_{I,F}\left[N_\chi(F) - N_\chi(I)\right]
        \int\du^{n_I}\Pi_I\dd^{n_F}\Pi_F(2\pi)^4\delta^4\left(p_I - p_F\right)
        \abs{\mathcal M_{I\to F}}^2\prod_{i\in I}f_i.
\end{equation}
Here $n_\chi$ denotes the number density of a dark species $\chi$, $I$ and $F$ index initial and final states, $N_\chi(S)$ denotes the number of $\chi$ particles in the state $S$, $\du\Pi_i=g_i\dd^3\bb p_i/(2\pi)^32E_i$, $\abs{\mathcal M_{I\to F}}^2$ is the spin-averaged squared matrix element, and $f_k$ is the phase space density of the species $k$. We assume Maxwell-Boltzmann statistics, and by conservation of comoving entropy density, we rewrite the left-hand side of \cref{eq:boltzmann} as $\dot n_\chi + 3Hn_\chi = S\dot Y_\chi$, where $S=(2\pi^2/45)g_{\star S}T^3$ is the entropy density and $Y_\chi\equiv n_\chi/S$. In turn, since $\dot T\approx -HT$, we have $S\dot Y_\chi \approx xHSY_\chi^\prime(x)$, where $x=\mu/T$ for any fixed mass $\mu$.

In freeze-in, one assumes that the phase space density of the dark species is always small, so that any initial state with $N_\chi(I)>0$ makes a negligible contribution in \cref{eq:boltzmann}. If all of the initial-state species are now in equilibrium, the phase space densities $f_i$ can be replaced with equilibrium distributions $e^{-E_i/T}$. Now \cref{eq:boltzmann} reads
\begin{equation}\label{eq:boltzmann-freeze-in}
    Y_\chi^\prime(x) = 
        \frac{1}{xHS}\sum_{I\not\ni\chi,F}N_\chi(F)
        \int\du^{n_I}\Pi_I\dd^{n_F}\Pi_F(2\pi)^4\delta^4\left(p_I - p_F\right)
        \abs{\mathcal M_{I\to F}}^2\exp\left(-xE_I/\mu\right).
\end{equation}

We will be interested in two types of processes: $1\to2$ decays and $2\to2$ scattering. In the $1\to 2$ case, with a process $i\to\chi f$, we set $\mu=m_i$, i.e., $x=m_i/T$. We recognize the decay width $\Gamma_{i\to \chi f}$ in \cref{eq:boltzmann-freeze-in}, which becomes
\begin{equation}\label{eq:freeze-in-decay}
    Y_\chi^\prime(x) = 
        \frac{1}{2\pi^2}\frac{g_im_i^3}{x^2HS}N_\chi(F)
        \Gamma_{i\to\chi f}K_1\left(x\right),
\end{equation}
where $K_1$ is a modified Bessel function of the second kind, and now $N_\chi(F)$ is either 1 or 2, depending on whether $f=\chi$. Substituting $H=1.66g_\star^{1/2}x^{-2}m_i^2\mpl^{-1}$, the total yield can now be computed by performing a one-dimensional integration of \cref{eq:freeze-in-decay} as
\begin{equation}
    Y_\chi(\infty) = 
        \frac{45N_\chi(F)g_i\mpl\Gamma_{i\to\chi f}}{1.66\times4\pi^4m_i^2}
        \int_{x_{\mathrm{min}}}^\infty\du x
        \frac{x^3K_1\left(x\right)}
            {g_\star^{1/2}g_{\star S}}.
\end{equation}
In particular, suppose that $f=\chi$, $m_\chi\ll m_i$, and $\abs{\mathcal M_{i\to\chi\chi}}^2 = \lambda^2$. If production mainly takes place during an epoch when $g_{\star}$ and $g_{\star S}$ are not changing rapidly, then we can estimate the yield as
\begin{equation}
    Y_\chi(\infty) \simeq 
        \frac{135N_\chi(F)g_i\mpl\lambda^2}
        {1.66\times8(2\pi)^{4}g_\star^{1/2}g_{\star S}m_i^3}
        \begin{cases}
            1 & x_{\mathrm{min}}\ll 1 \\
            \frac13\sqrt{\frac{2}{\pi}}\,
            x_{\mathrm{min}}^{5/2}
                \exp\left(-x_{\mathrm{min}}\right)
                & x_{\mathrm{min}}\gg 1.
        \end{cases}
\end{equation}
Similarly, if the abundance of $\chi$ is set by $2\to2$ processes of the form $ij\to\chi f$, then the integrals over the final-state phase space produce the cross section $\sigma_{ij\to\chi f}$, and \cref{eq:boltzmann-freeze-in} becomes
\begin{equation}\label{eq:yield-estimate-decay}
    Y_\chi^\prime(x) = 
        \frac{N_\chi(F)g_ig_j}{xHS}
        \int\frac{\du^3\bb p_i}{(2\pi)^3}\frac{\du^3\bb p_j}{(2\pi)^3}
            \,\sigma v\exp\left(-xE_i/\mu\right)\exp\left(-xE_j/\mu\right).
\end{equation}
These remaining integrals can be reduced to a single 1D integral, following e.g. \cite{Gondolo:1990dk}. Integrating in $x$, the yield is then
\begin{multline}
    Y_\chi(\infty) = \frac{\mu N_\chi(F)g_ig_j}{2(2\pi)^4}
    \int_{x_{\mathrm{min}}}^\infty
        \frac{\du x}{x^2HS}
    \int_{s_{\mathrm{min}}}^\infty\du s\;\sigma(s)\,
        r_-r_+\times\\
        \times\Bigl\{\frac{m_+m_-}{s}\left(\frac{\mu}{x}+\sqrt s\right)
            \exp\left(-x\sqrt s/\mu\right)
        + \frac{r_-r_+}{\sqrt s}K_1\left(x\sqrt s/\mu\right)
    \Bigr\}
    ,
\end{multline}
where $m_\pm = \left|m_i\pm m_j\right|$, $r_\pm = \left(s-m_\pm^2\right)^{1/2}$, and $s_{\mathrm{min}} = \min(m_i+m_j,\,m_\chi+m_f)^2$. As in the $1\to2$ case, we can estimate the yield analytically for a process $ii\to\chi\chi$ when $m_i\ll m_\chi$ and the evolution of $g_\star$ and $g_{\star S}$ is negligible. If $\abs{\mathcal M_{ii\to\chi\chi}}^2 = \lambda^2$, then the result is
\begin{equation}\label{eq:yield-estimate-scattering-s-independent}
    Y_\chi(\infty) \simeq \frac{45 N_\chi(F)g_i^2\mpl\lambda^2}
        {1.66\times512\pi^5g_{\star}^{1/2}g_{\star S}m_i}
        \begin{cases}
            (3\pi/8)m_i/m_\chi & x_{\mathrm{min}} \ll 1 \\
            x_{\mathrm{min}}\exp\left(-2x_{\mathrm{min}}m_\chi/m_i\right)
                & x_{\mathrm{min}} \gg 1,
        \end{cases}
\end{equation}
where $x_{\mathrm{min}}=m_i / T_{\mathrm{max}}$. The analogous expression for $m_\chi\ll m_i$ is obtained by interchanging $m_i$ and $m_\chi$ and taking $\mu=m_\chi$ (i.e., $x_{\mathrm{min}} = m_\chi / T_{\mathrm{max}}$). However, in our model, $2\to2$ processes are driven by effective four-point vertices suppressed by a scale $\LNP$, so we should instead set $\abs{\mathcal M_{ii\to\chi\chi}}^2 = s/\LNP^2$. In this case, the result is
\begin{equation}\label{eq:yield-estimate-scattering-s-dependent}
    Y_\chi(\infty) \simeq \frac{45 N_\chi(F)g_i^2\mpl m_\chi^2}
        {1.66\times128\pi^4g_{\star}^{1/2}g_{\star S}m_i\LNP^2}
        \begin{cases}
            \frac{8}{\pi}\left(m_\chi / m_i\right)^{-2}x_{\mathrm{min}}^{-1}
                & x_{\mathrm{min}} \ll 1 \\
            x_{\mathrm{min}}\exp\left(-2x_{\mathrm{min}}m_\chi/m_i\right)
                & x_{\mathrm{min}} \gg 1.
        \end{cases}
\end{equation}
This demonstrates a key difference between standard freeze-in and UV freeze-in: A naive extrapolation of the production rate to arbitrarily high temperatures (small $x_{\mathrm{min}}$) diverges. Of course, one should not expect to accurately compute the production rate in the effective theory at $T\gg\LNP$. But even so, if $\LNP\gg T_{\mathrm{max}}\gg \max\{m_\chi, m_i\}$, then production can be dominated by $2\to2$ processes, whereas $1\to 2$ decays typically dominate in standard freeze-in. In our case, $m_\chi$ and $m_i$ are MeV scale, while $\LNP\gtrsim\SI{e6}{\giga\electronvolt}$. Thus, production by $2\to2$ processes at high temperatures is potentially very significant.

Using the approximate forms of the yield derived above together with the DM abundance today $Y_\chi(\infty)\approx2\times10^{-6}(m_\chi/\SI{}{\mega\electronvolt})$, we can estimate the ranges of parameters which account for all of DM---or, at least, those which do not overclose the Universe. If DM in our model is produced dominantly by quark annihilation via an interaction of the form $\Ldd^{-1}d(i\gamma_5)\bar dSP$, then the only important parameters are $\Ldd$ and $x_{\mathrm{min}}$. Note that if this is the only interaction at work, there is no contribution from decays.

First, suppose that $x_{\mathrm{min}}\ll 1$. Then the scale $\Ldd$ must satisfy
\begin{equation}
    \Ldd \gtrsim
        \left(
            \frac{\left.g_{\star}\right|_{\TRH}}{100}
        \right)^{-3/4}
        \left(\frac{\TRH}{\SI{}{\giga\electronvolt}}\right)^{1/2}
        \SI{3e10}{\giga\electronvolt}
    .
\end{equation}
Per the analysis in \cref{sec:KOTO}, this is too large to account for the KOTO excess, and this estimate accounts for only one production channel! In particular, if $\TRH>\Ldd$, DM is dramatically overproduced. At the very least, one requires $\TRH\lesssim\SI{100}{\mega\electronvolt}$, where the approximations made for this estimate are no longer trustworthy. However, suppose instead that reheating indeed takes place near the MeV scale, so that $x_{\mathrm{min}}\gg 1$. Then the situation is quite different: Neglecting the difference between $m_S$ and $m_P$, we have
\begin{equation}\label{eq:scale-bound-MeV-reheating}
    \Ldd \gtrsim
        \left(\frac{\left.g_{\star}\right|_{\TRH}}{10}\right)^{-3/4}
        \left(\frac{\TRH}{\SI{10}{\mega\electronvolt}}\right)
        \exp\left[-\left(
            \frac{m_S}{\TRH}-30\right)
        \right]
        \SI{300}{\giga\electronvolt}
    .
\end{equation}
This bound poses no obstacle to accounting for the KOTO excess. When combined, these two estimates naively suggest that our model can account for all of DM if reheating takes place between \SI{100}{\mega\electronvolt} and \SI{10}{\mega\electronvolt}. While the scale of reheating is often assumed to be much higher, the strongest observational lower bound on the reheating temperature is, in fact, only $\TRH\gtrsim\SI{5}{\mega\electronvolt}$ \cite{Hannestad:2004px,deSalas:2015glj}. There is no particularly strong motivation for a very high reheating temperature, and certainly nothing inconsistent about reheating taking place at \SI{10}{\mega\electronvolt}.

However, production at such low temperatures introduces a new complication: Our simplistic estimates above have presumed not only that production is dominated by $2\to 2$ processes, but also that the initial state consists of free quarks. If $\TRH<\SI{100}{\mega\electronvolt}$, then quarks are confined into hadrons during the entire production period. One must then modify the effective couplings to account for hadronic scattering, and since the initial and final states are all (pseudo)scalars, the matrix elements no longer carry any $s$ dependence. Additionally, since single hadrons can now decay to $S$ and $P$, hadronic decays can dominate the relic abundance and must be included in the calculation of the yield.

In the following section, we treat these issues in detail and calculate the relic density numerically.

\subsection{Determining the reheating temperature}
Our estimates in the previous section suggest that $P$ can be produced nonthermally, and can account for all of DM if the initial temperature of the SM bath is between \SI{100}{\mega\electronvolt} and \SI{10}{\mega\electronvolt}. We now refine our estimate of the yield to account for confinement and hadronic decays, and then numerically compute the yield to establish the required reheating temperature in our model.

At $T\lesssim\SI{200}{\mega\electronvolt}$, quarks are confined into hadrons, and the effective interactions of the hadrons with $S$ and $P$ are well described by chiral perturbation theory (chiPT). The effective couplings of hadrons to $S$ and $P$ are built from a combination of the new physics scales and QCD parameters. Since the couplings in the quark-level effective Lagrangian are proportional to $\LNP^{-1}$, and the hadron-level $1\to 2$ coupling must have mass dimension one, the latter must be of order $\Lchi^2/\LNP$, where $\Lchi$ is some scale associated with low-energy QCD. Similarly, in the $2\to2$ case, the hadron-level coupling should have the form $\Lchi^\prime/\LNP$. As we will see momentarily, $\Lchi^{(\prime)}$ is a combination of two constants, $f_\pi\approx\SI{92}{\mega\electronvolt}$ and $B_0\approx\SI{2666}{\mega\electronvolt}$. To determine the couplings explicitly, we match our effective quark-level Lagrangian onto the chiPT Lagrangian following \cite{Pich:1995bw,Gasser:1984gg}. Our application of this method to light scalars is also similar to the treatment in section 3.1 of \cite{Ziegler:2020ize}.

The interactions of QCD degrees of freedom with our light scalars can be written as the couplings of quarks to external currents $s$ and $p$, respectively a scalar and pseudoscalar. These take the form
\begin{equation}
    \mathcal L_{\mathrm{QCD}}[s,p] =
        -\bar{\bb q}\left( s(x) - i\gamma_5p(x)\right)\bb q.
\end{equation}
Interactions of hadrons with these currents enter the chiPT Lagrangian via the current $\chi = 2B_0(s+ip)$. At lowest order, we have
\begin{equation}\label{eq:L2-interactions}
    \mathcal L_2 \supset
        \frac{f_\pi^2}{4}\tr\left(\chi U^\dagger + U\chi^\dagger\right),
    \qquad
    U = \exp\left(\frac{i\sqrt2}{f_\pi}\;\Phi\right),
\end{equation}
where $\Phi$ is the pseudo-Nambu--Goldstone boson (PNGB) matrix \cite[see, e.g.,][]{Pich:1995bw}. Now consider a quark-level interaction of the form
\begin{equation}
    \mathcal L \supset
        \frac12\bar q_i\left(
            \gq{ij}{\mathcal O_S} - i\gqp{ij}{\mathcal O_S}\gamma_5
        \right)q_j\mathcal O_S
        +
        \frac i2\bar q_i\left(
            \gq{ij}{\mathcal O_P} - i\gqp{ij}{\mathcal O_P}\gamma_5
        \right)q_j\mathcal O_P
        +
        \mathrm{H.c.}
        ,
\end{equation}
where $\mathcal O_S$ is a scalar ($CP$ even) and $\mathcal O_P$ is a pseudoscalar ($CP$ odd). We can then identify
\begin{align}
    &s_{ij} = -\frac12\left(
        \gq{ij}{\mathcal O_S} + \gqc{ji}{\mathcal O_S}
    \right)\mathcal O_S - \frac 12\left(
        \gq{ij}{\mathcal O_P} - \gqc{ji}{\mathcal O_P}
    \right)\mathcal O_P,
    \\
    &p_{ij} = -\frac i2\left(
        \gqp{ij}{\mathcal O_S} - \gqpc{ji}{\mathcal O_S}
    \right)\mathcal O_S - \frac i2\left(
        \gqp{ij}{\mathcal O_P} + \gqpc{ji}{\mathcal O_P}
    \right)\mathcal O_P.
\end{align}
Substituting these expressions into \cref{eq:L2-interactions} with $\mathcal O_S = S^2,\, P^2$, and $\mathcal O_P = SP$ gives the interactions of $S$ and $P$ with the PNGBs. For instance, the interactions of $S$ and $P$ with $\pi^0$ are specified by
\begin{multline}
    \label{eq:pi0-interactions}
    \mathcal L_2 \supset B_0f_\pi\pi^0\left(
        SP\im\gq{dd}{SP} - S^2\im\gqp{dd}{S^2} - P^2\im\gqp{dd}{P^2}
    \right) \\
    - \frac12B_0(\pi^0)^2\left(
        SP\re\gqp{dd}{SP} - S^2\re\gqp{dd}{S^2} - P^2\re\gqp{dd}{P^2}
    \right) + \dotsb,
\end{multline}
where the ellipsis denotes a series of higher-dimensional operators. We include all terms up to second order in the PNGB fields in our analysis, and the form of the hadron-level Lagrangian is as expected from dimensional analysis. Note that it is essential to consider complex-valued $\gq{ij}{}$ and $\gqp{ij}{}$, without which some interactions will vanish.

We can now determine the reheating temperature required to produce the observed DM density as a function of our model parameters.
First, using the normalization factors as they appear in \cref{eq:pi0-interactions}, we can now estimate the relative significance of decays and scattering, starting with \cref{eq:yield-estimate-decay,eq:yield-estimate-scattering-s-independent}. Assuming that all dimensionless couplings are $\mathcal O(1)$, we set the coupling $\lambda$ for three-point vertices equal to $B_0f_\pi/\LNP$, and we set the coupling for four-point vertices to $B_0/\LNP$. In this regime, we typically have $m_i\gg \max\{m_P,\TRH\}$, and in this limit,
\begin{equation}
    \frac{Y_{P}^{1\to 2}(\infty)}
         {Y_{P}^{2\to 2}(\infty)} \simeq
    32\left(\frac{f_\pi}{m_i}\right)^2\begin{cases}
        1 & m_P \ll \TRH \ll m_i \\
        \frac{3\pi}{8}(\TRH/m_i)\exp\left(2m_i/\TRH\right) & \TRH \ll m_P \ll m_i.
    \end{cases}
\end{equation}
Our parameter space includes $\SI{1}{\mega\electronvolt} \lesssim m_P \lesssim \SI{200}{\mega\electronvolt}$, so the ratio above can be large or $\mathcal O(1)$ depending on the choice of the $P$ mass, but it is never small. Note, however, that increasing $m_P$ can also close certain decay channels. In particular, if there exist interactions allowing the decay $\pi^0 \to PP$, this channel naively dominates production at low temperatures, but is closed for $2m_P > m_{\pi^0}$.

Since decays dominate in most of the parameter space, we can make a first estimate of the yield by considering only production via $K_L \to SP$, the same decay process which is necessary to account for the KOTO excess. Neglecting the distinction between $m_S$ and $m_P$, the yield is
\begin{equation}
    Y_P^{K_L\to SP}(\infty) \simeq
    \frac{45}{1.66\times4(2\pi)^{9/2}g_{\star}^{1/2}g_{\star S}}
    \left(\frac{Bf_\pi}{m_K\Lsd}\right)^2
    \frac{\mpl}{\TRH}
    \exp\left(-2m_S/\TRH\right),
\end{equation}
and the resulting upper bound on $\Lsd$ is
\begin{equation}
    \Lsd \gtrsim
        \left(\frac{\left.g_{\star}\right|_{\TRH}}{10}\right)^{-3/4}
        \left(\frac{\TRH}{\SI{15}{\mega\electronvolt}}\right)^{1/2}
        \exp\left[
            -\left(\frac{m_S}{\TRH} - 20\right)
        \right] \SI{5e6}{\giga\electronvolt}.
\end{equation}
For the typical parameter values selected above, this upper bound is toward the lower edge of our parameter space of interest for the KOTO excess. Thus, although hadronic decays significantly enhance production relative to the prediction of \cref{eq:scale-bound-MeV-reheating}, this channel on its own does not pose an obstacle to accounting for the KOTO excess.

\begin{figure}\centering
    \includegraphics[width=\textwidth]{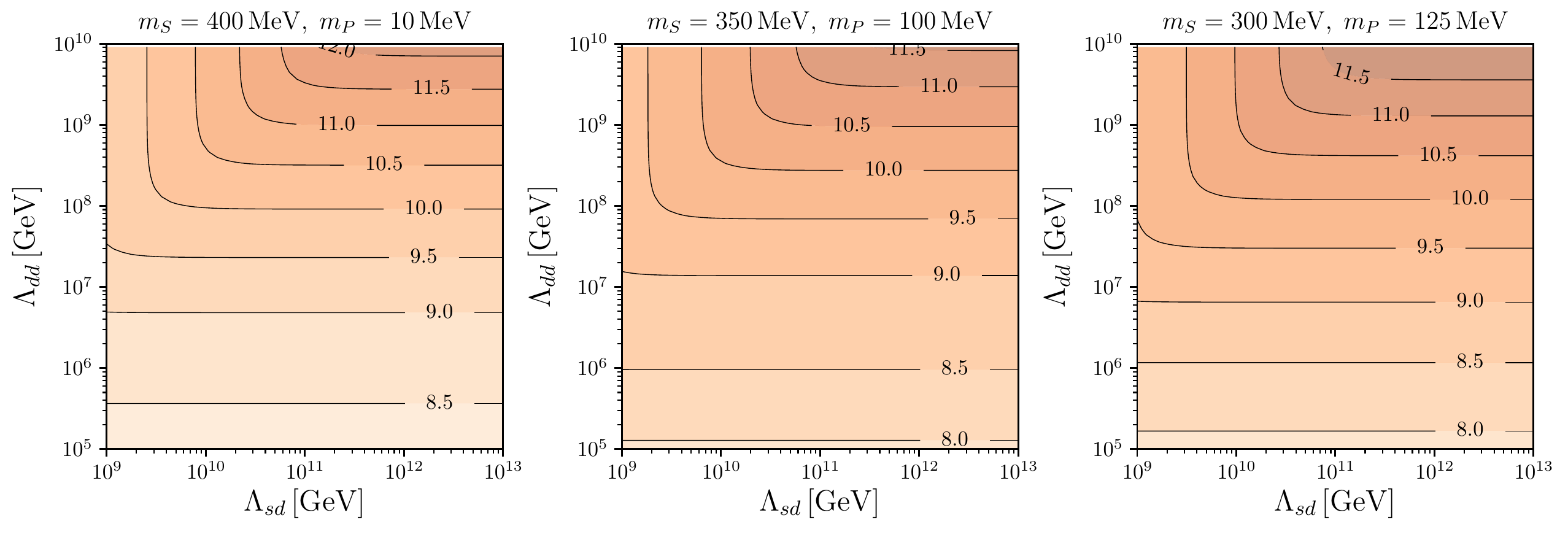}
    \caption{Reheating temperature in MeV to produce the observed DM relic density, including all production channels with no DM in the initial state. The couplings $\gq{sd}{SP}$ and $\gq{dd}{SP}$ are taken to be purely imaginary, while all other couplings are set to zero, corresponding to the minimal scenario to account for the KOTO excess. In the leftmost panel (BM1), all decay channels are open. In the middle panel (BM2), $S\to 3P$ is kinematically closed, so there are no number-changing interactions in the dark sector: $S$ decays via $S\to\pi^0P$. In the rightmost panel (BM3), $S\to 3P$ and $\pi^0\to PP$ are both closed, so there is no contribution to the relic density from $\pi^0$ decays.}
    \label{fig:reheating-temperature}
\end{figure}

However, in general, it is necessary to numerically evaluate the yield to determine the extent of the viable parameter space, and, in particular, to identify the reheating temperature that produces the observed relic density at each parameter point. The resulting reheating temperatures are shown in \cref{fig:reheating-temperature}, and are of order \SI{10}{\mega\electronvolt} throughout the parameter space of interest. The required reheating temperature is mainly controlled by the smaller of $\Lsd$ and $\Ldd$, with a slight bias toward $\Lsd$, since production by $\eta$ decays is suppressed compared to production by $K^0$ decays due to their relative masses. Note that all couplings except for $\gq{sd}{SP}$ and $\gq{dd}{SP}$ are neglected in \cref{fig:reheating-temperature}, so, in particular, $\pi^0\to PP$ does not contribute to the relic density even when $2m_P < m_{\pi^0}$. If we suppose that all of the couplings in the effective theory are of similar order, the viable parameter space can change significantly.

\begin{figure}\centering
    \includegraphics[width=\textwidth]{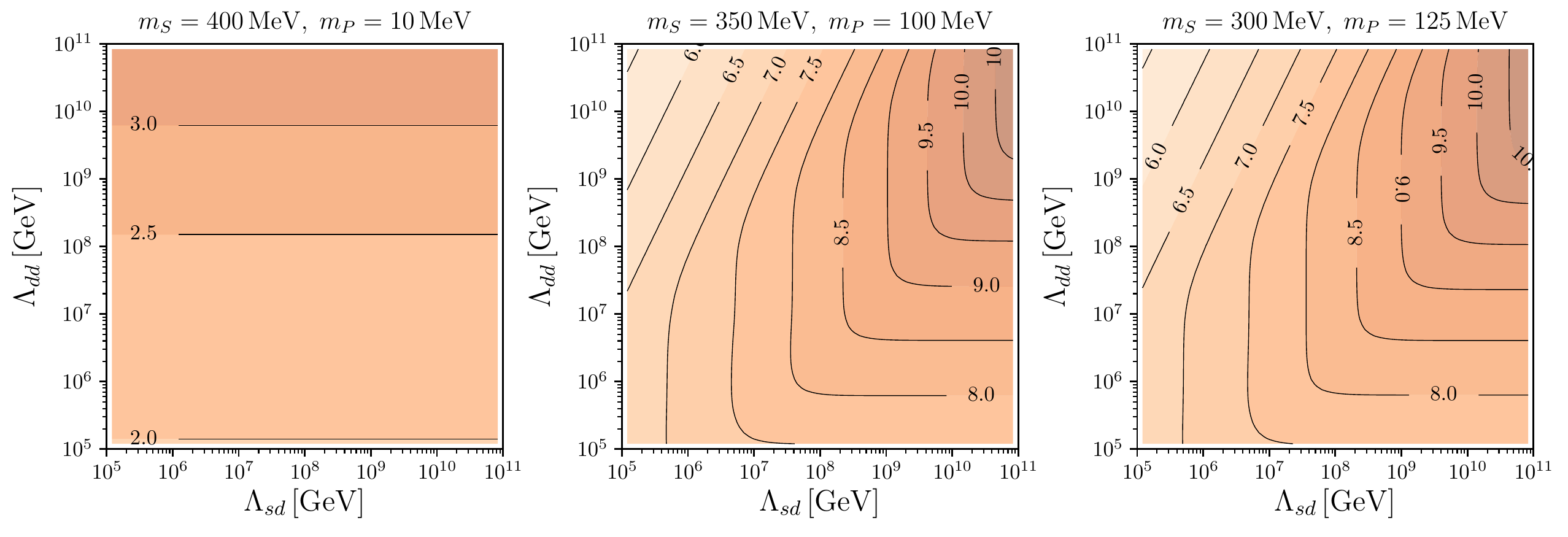}
    \caption{Reheating temperature (in MeV) to produce the observed DM relic density, including all production channels with no DM in the initial state, as in \cref{fig:reheating-temperature}. Here it is assumed that $S^2$, $P^2$, and $SP$ couple equally to light quark bilinears, and that $\gq{sd}{\mathcal O}$ is the geometric mean of $\gq{ss}{\mathcal O}$ and $\gq{dd}{\mathcal O}$. The real and imaginary parts of all couplings are taken to be equal. In the first panel (BM1), all decay channels are open, and production is dominated by $\pi^0$ decays. In the middle panel (BM2), $\pi^0\to PP$ is closed, but $S\to 3P$ is still open. In the rightmost panel (BM3), both $\pi^0\to PP$ and $S\to 3P$ are closed, so $S$ decays only via $S\to\pi^0P$. In the leftmost panel, since production is dominated by $\pi^0\to PP$, the relic abundance is controlled exclusively by $\Ldd$. In this case, the required reheating temperatures are observationally inviable throughout the parameter space. In the other two panels, production is dominated by $K^0$ and $\eta$ decays, and their relative importance depends on $\Lsd$ and $\Ldd$.}
    \label{fig:reheating-temperature-all-couplings}
\end{figure}

We can estimate this effect by taking $\gq{q_1q_2}{S^2}=\gq{q_1q_2}{SP}=\gq{q_1q_2}{P^2}$ and setting $\gq{sd}{\mathcal O} = \left(\gq{ss}{\mathcal O}\gq{dd}{\mathcal O}\right)^{1/2}$ to fix $\gq{ss}{\mathcal O}$. The resulting reheating temperatures are shown in \cref{fig:reheating-temperature-all-couplings}. With these choices for the couplings, our two benchmark points with $m_P=\SI{10}{\mega\electronvolt}$ are incompatible with freeze-in as a production mechanism, since the required reheating temperature is below observational bounds throughout the relevant parameter space. This is due to the open $\pi^0\to PP$ decay, which is kinematically closed for the other two benchmark points with $m_P=\SI{100}{\mega\electronvolt}$ and $m_P=\SI{125}{\mega\electronvolt}$. For these points, the required reheating temperature is again of order \SI{10}{\mega\electronvolt} throughout the relevant parameter space.
At the top left of the corresponding panel of \cref{fig:reheating-temperature-all-couplings}, the required reheating temperature \emph{decreases} with increasing $\Ldd$. This is just because of our assumption that $\gq{sd}{\mathcal O}$ is the geometric mean of $\gq{dd}{\mathcal O}$ and $\gq{ss}{\mathcal O}$: Increasing $\Ldd$ corresponds to decreasing $\gq{dd}{\mathcal O}$, so if $\gq{sd}{\mathcal O}$ is held fixed, then $\gq{ss}{\mathcal O}$ must increase to compensate. This increases the relic density, forcing a lower reheating temperature.

Finally, we note that the reheating temperatures shown in \cref{fig:reheating-temperature,fig:reheating-temperature-all-couplings} are potentially imprecise, and should be viewed as lower bounds. Our calculation of the yield assumes that all of the initial-state species are thermalized, but the mesons freeze out at temperatures of the same order considered here. In particular, $\pi^0$, $K^0$, and $\eta$ freeze out at  \SI{3}{\mega\electronvolt}, \SI{10.5}{\mega\electronvolt}, and \SI{11.6}{\mega\electronvolt}, respectively. In a scenario with a high reheating temperature, this concern would be less significant: The mesons would have a thermal distribution at early times, so as long as DM production is not dominated by temperatures well below the mesons' freeze-out temperatures, the effect should be small. However, we are speculating that the reheating temperature itself is lower than e.g. the kaon freeze-out temperature in parts of our parameter space, in which case the kaons may never be populated with anything resembling a thermal distribution. It is thus possible that \cref{eq:boltzmann-freeze-in} overestimates the DM relic abundance.

This does not have a significant effect on our qualitative results: We can safely predict that DM is overproduced if $\TRH\gtrsim\SI{15}{\mega\electronvolt}$, in which case all of the relevant mesons are thermalized, so this is an upper bound on $\TRH$. Likewise, we can see that DM would be underproduced for $\TRH$ below a particular value even if the mesons have their equilibrium number densities.\footnote{Since production is dominated by decays, the DM relic abundance is mainly determined only by the number density of the parent mesons, and is fairly insensitive to other details of the phase space distribution.} This lower threshold is $\mathcal O(\SI{7}{\mega\electronvolt})$ if $\pi^0\to PP$ is forbidden, and $\mathcal O(\SI{2}{\mega\electronvolt})$ if it is not. The only qualitative importance of out-of-equilibrium effects is that it may be possible to construct a cosmologically viable model in which DM is not overproduced even if $\pi^0\to PP$ is open. However, such a model would depend on the details of reheating, and this analysis lies beyond the scope of this work.

\section{Discussion} \label{sec:discussion}
In the foregoing sections, we have introduced a model to account for the KOTO excess and explored the cosmological effects. We now discuss the implications of our results and future experimental prospects.

If the KOTO excess is interpreted at face value, this suggests apparent violation of the GN bound. As has been discussed by several authors \cite{Fuyuto:2014cya,Kitahara:2019lws,Egana-Ugrinovic:2019wzj,Dev:2019hho,Jho:2020jsa,Liu:2020qgx,He:2020jzn,Ziegler:2020ize,Liao:2020boe,Gori:2020xvq,Hostert:2020gou,Datta:2020auq}, such a signal at KOTO can be mimicked by a decay of the form $K_L\to \pi^0 X$, where $X$ denotes one or more invisible species. In contrast to most studies, we focus on a new physics scenario where the decay $K_L \to \pi^0 ~\text{inv.}$ is realized through a sequence of two-body decays $K_L \to SP \to \pi^0 PP$, where $S$ and $P$ are light neutral scalar particles. Similar scenarios were also studied in~\cite{Hostert:2020gou} where the light particles interact with the SM through a vector or scalar portal. Here we instead analyze a setup where $S$ and $P$ are coupled to the SM through effective operators at a characteristic new physics scale of $\LNP \sim 10^6$--\SI{e9}{\giga\electronvolt}. We have stabilized $P$ with a $Z_2$ symmetry under which SM species are even and our new species are odd, and we have entertained the possibility of other interactions consistent with such $Z_2$ invariance, including an $SP^3$ term that could mediate the decay of $S\to 3P$. Our effective theory is readily UV completed by, e.g., very heavy vectorlike quarks or a TeV-scale inert Higgs doublet.
Such UV completions can realize a minimal case in which only interactions between SM quarks and $SP$ are present at low energies, as well as more generic cases that include interactions with $S^2$ and $P^2$.

If the KOTO excess persists, the GN bound heavily constrains new physics interpretations. A model of the type we consider, with new light scalars, is one of the simplest and most elegant solutions. Since the scale $\LNP\sim10^6\text{--}\SI{e9}{\giga\electronvolt}$ indicated by the KOTO excess is so large, most other experiments are not substantially constraining (with the notable exception of beam-dump experiments, to which we will return shortly). In particular, in our scenario, there is a large region of parameter space which can account for the KOTO excess while still unconstrained by other rare meson decays. However, it is important to consider astrophysical constraints. Supernova cooling limits can potentially rule out lower $P$ masses: As discussed in \cref{sec:supernova}, supernova temperatures are high enough, at 10's of MeV, to probe the lightest $S$ and $P$ masses that we consider in \cref{fig:masses}. These constraints are most significant for $\Ldd\lesssim\SI{e6}{\giga\electronvolt}$, and it is important to note that establishing firm constraints from supernova cooling requires a much more detailed analysis beyond the scope of this work. However, the simplistic expectation is that $P$ masses of $\mathcal O(\SI{10}{\mega\electronvolt})$ and below are disfavored, making our scenario easier to test.

Since the KOTO excess motivates the introduction of new feebly coupled particles, it is natural to speculate that these new species might contribute to cosmological DM, and indeed, we have shown that $S$ and $P$ can constitute all of DM even in the most minimal scenarios needed to explain the KOTO signal. Nevertheless, this comes at a cost: In the absence of additional interactions, there is no mechanism to reduce the DM abundance, and cosmological reheating must take place at very late times, at a temperature of order \SI{10}{\mega\electronvolt}. This requirement should be interpreted as a cosmological constraint on our model and similar models accounting for the KOTO excess. The scale of the preferred reheating temperature originates mainly from the masses of the new scalars: Since the DM abundance is exponentially suppressed in $m_{\mathrm{DM}} / \TRH$, the required reheating temperature depends only logarithmically on the couplings and other scales of new physics.

Such a thermal history is necessary because the effective coupling lies in an intermediate regime: It is too small for freeze-out to deplete the DM abundance but large enough that UV freeze-in generically overproduces DM. Thus, an additional feature is needed to prevent overproduction. The simplest mechanism to accomplish this, without any modification to the model, is to make a judicious choice of the reheating temperature. Since we are working with an effective theory, the DM relic density is inherently sensitive to the reheating temperature; indeed, if $\TRH\gtrsim\LNP$, we cannot consistently calculate the relic density, but only bound it below. Thus, since $\TRH$ is necessarily a parameter of our model, $\TRH\sim\SI{10}{\mega\electronvolt}$ is as natural as any other choice. As we have discussed, observational constraints are ineffective at temperatures above $\sim\SI{5}{\mega\electronvolt}$.

We note that in principle low-temperature reheating might leave an imprint on early Universe probes such as BBN and the cosmic microwave background (CMB). Unfortunately, such potential signals are highly model dependent. Specifically, low-reheating temperature scenarios have been shown in the literature to impart a significant effect on the synthesis of light elements, primarily via (i) modifications to the Hubble rate around BBN by changing the energy density of both relativistic and matter species; (ii) changing the momentum distribution of electron-flavor neutrinos, which directly enters charged current interactions, in turn governing the neutron-proton chemical equilibrium; and (iii) by entropy exchange that can affect the ratio of neutrino to photon temperature, which in turn is testable with CMB data.

Previous studies (see, e.g., \cite{Hasegawa:2020ctq} and references therein) relied on simple assumptions such as a single massive matter species driving reheating, and decaying primarily into neutrinos \cite{deSalas:2015glj}, or electromagnetically interacting species \cite{Hannestad:2004px}, or hadrons \cite{Hasegawa:2019jsa}. Generally, testable effects arise for $\TRH\lesssim 5$ MeV, implying that no signal is expected for the scenario discussed here, where $\TRH\gtrsim 10$ MeV. However, it is important to point out that the reheating scenario might include features that could manifest themselves when more stringent probes of CMB become available in the future \cite{CMBS4}. For instance, the field driving reheating might actually be an \textit{ensemble} of fields, with different masses; the $S$ and $P$ particles might be directly produced in the decay of the field(s) driving reheating, changing the predictions for $\TRH$ made above; or new physics in the neutrino sector could make reheating temperatures in the 10's of MeV visible once constraints on $N_{\mathrm{eff}}$ significantly improve.

There are other mechanisms which prevent the overproduction of DM without requiring a particular temperature for reheating. One possibility is to add an interaction with the SM to restore freeze-out as a viable thermal history, as we discussed briefly in the context of a neutrino portal. This would be a heartening scenario: Reheating can still take place at a very high temperature, and the coupling to leptons might allow for additional experimental probes. However, there are several other possibilities. In particular, it is possible that the DM abundance is depleted by additional interactions within the dark sector. This is not possible in our effective theory, but one can consider extensions which keep the DM in thermal equilibrium long after decoupling from the SM bath, or which allow other number-changing processes at a sufficient rate to allow for freeze-out at high temperatures. We emphasize again that our results imply cosmological constraints on models of the KOTO excess: Cosmology requires either a restricted range of reheating temperatures or additional features of the low-energy theory, regardless of what fraction of cosmological DM is composed of $P$.

Of course, one can also consider constraints which only apply if $P$ makes up a significant fraction of DM. The simplest of these is the Lyman-$\alpha$ constraint on warm DM \cite{Viel:2005qj}, which requires the $P$ population to be nonrelativistic at temperatures of $\mathcal O(\SI{}{\kilo\electronvolt})$. If $P$ is produced nonthermally via decays at \SI{10}{\mega\electronvolt}, typical energies will be of order the masses of the parent states, i.e., $\mathcal O(\SI{100}{\mega\electronvolt})$. Thus, in order for $P$ to be nonrelativistic when $T_\gamma\sim\SI{}{\kilo\electronvolt}$, we require that $m_P\gtrsim\SI{10}{\kilo\electronvolt}$. This is a somewhat weaker bound than one expects from supernovae, but it is not subject to the complicated physics involved in such constraints.

The annihilation cross section into visible states is much too small ($\sim\SI{e-50}{\centi\meter^2}$) for indirect detection to be viable, nor is there any significant self-interaction in the dark sector. However, the scattering cross section with nuclei could be as large as $\sim 0.1$ pb, and thus potentially within reach of future, planned experimental sensitivity for sub-GeV direct DM searches. It is thus possible (albeit not guaranteed) that future experiments will probe such signatures associated with our model---particularly direct detection---but it is important to note that in the minimal scenario for the KOTO excess, these signatures are substantially suppressed even compared to the generic expectation. This is because the KOTO excess only requires SM interactions with the current $SP$, and not $PP$. Since any DM accounted for by our model is composed entirely of $P$, this means that any diagrams contributing to indirect detection must be suppressed by $\LNP^{-4}$. Moreover, at lowest order, direct detection is only sensitive to the inelastic scattering process $NP\to NS$, which is kinematically prohibited for nonrelativistic DM. It is thus challenging to conclusively establish that $P$ makes up cosmological DM through direct observational means.

\begin{figure}[tb]
\centering
\includegraphics[width=0.48\textwidth]{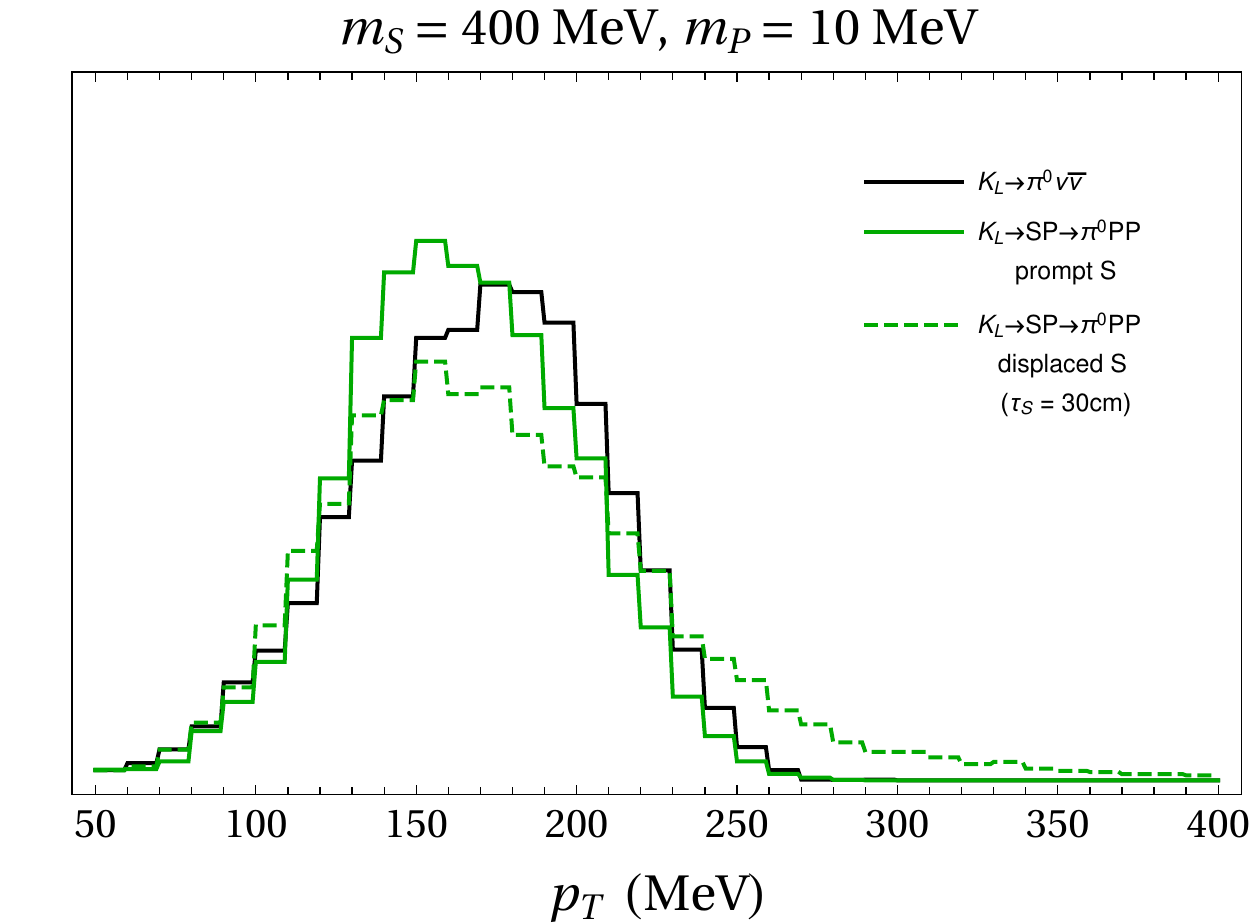} ~~~
\includegraphics[width=0.48\textwidth]{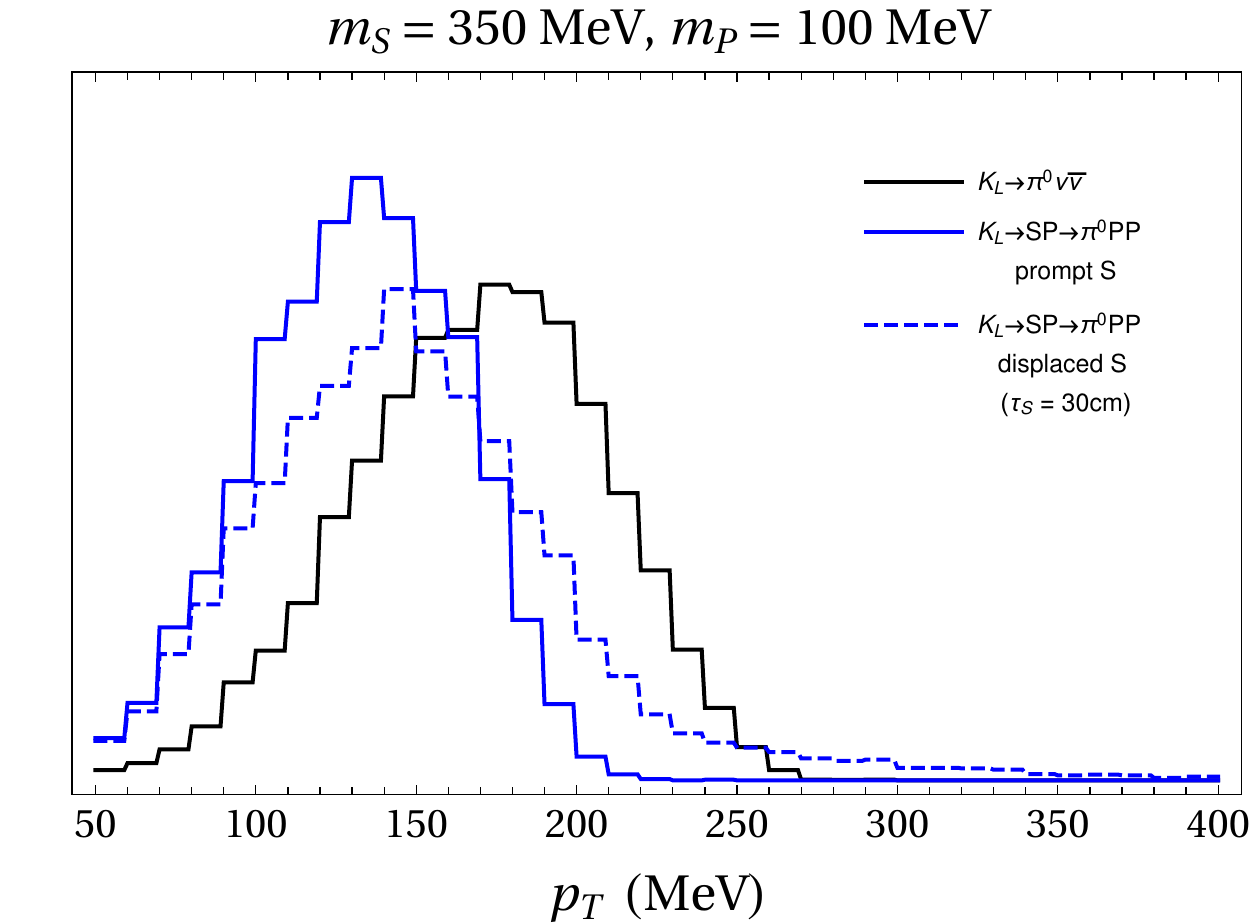} \\[12pt]
\includegraphics[width=0.48\textwidth]{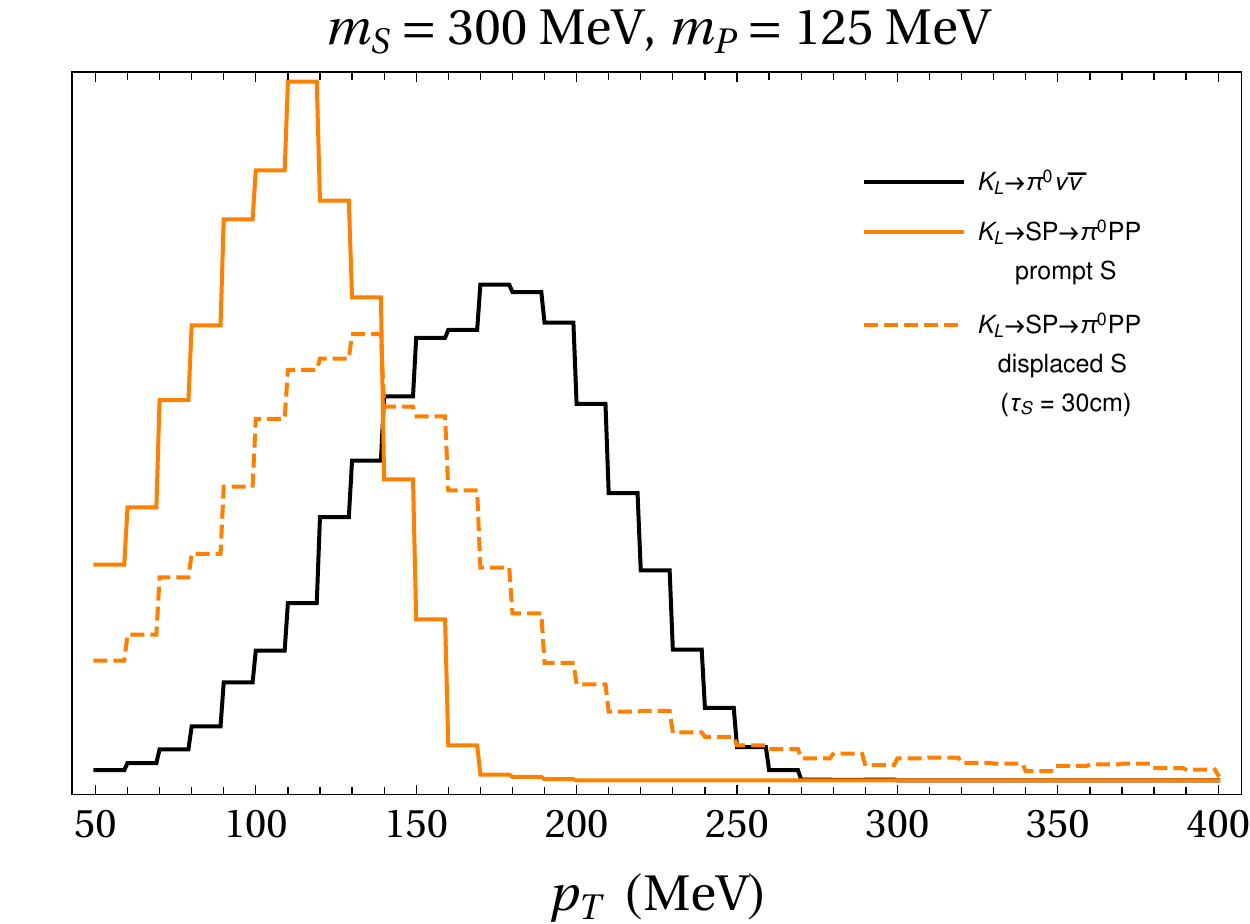} ~~~
\includegraphics[width=0.48\textwidth]{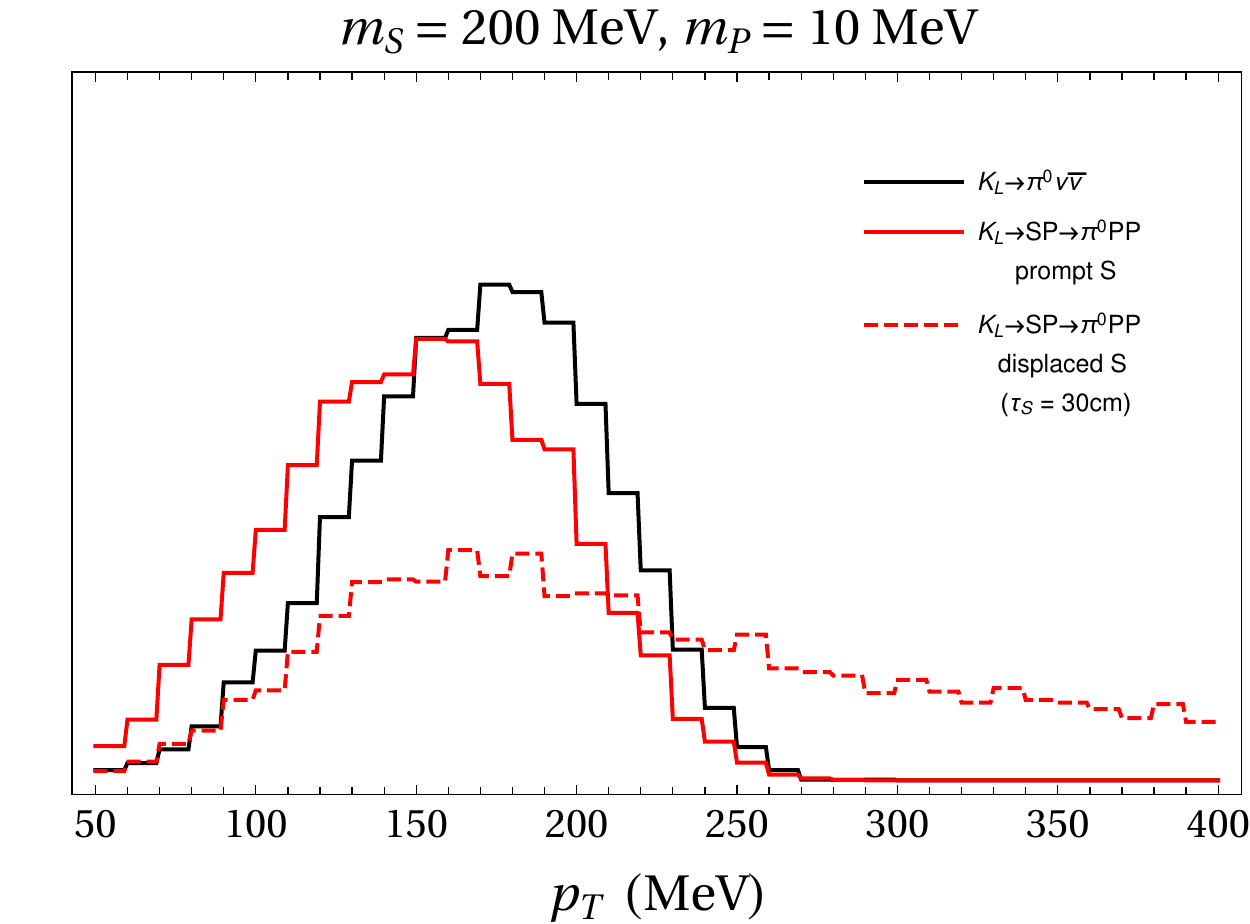} 
\caption{Pion $p_T$ distributions for the $K_L \to \pi^0 \nu\bar\nu$ decay and the $K_L \to SP \to \pi^0 PP$ decay in our benchmark points. The distributions are shown both for prompt $S$ decays and $S$ decays with a lifetime of \SI{30}{\centi\meter}.}
\label{fig:pion-pT}
\end{figure}

However, it is potentially much easier to determine whether a model like ours accounts for the KOTO excess. If the excess persists at its present size, then as KOTO reaches its design sensitivity, hundreds of events will be observed. With a sample of this size, it is possible to distinguish our model from SM three-body decays kinematically in much of our parameter space, simply by measuring the pion's transverse momentum. In \cref{fig:pion-pT}, we show the transverse momentum distributions expected at KOTO in the SM and in our model. By sampling from these distributions and applying the Kolmogorov--Smirnov test, we find that the $p_T$ distribution in our model can be distinguished from the SM three-body decay at $5\sigma$ with $\mathcal O(100)$ events in much of our parameter space. Sensitivity is lost when $m_P$ is small and $m_S \sim m_{K_L}$, and the distributions may also be too close to distinguish at smaller $m_S$ if the $S$ lifetime is shorter than $\mathcal O(\SI{10}{\centi\meter})$. Still, there are good prospects for making such a determination within the next several years, as KOTO continues to collect data.

There are also discovery prospects for $S$ particles with meter- and centimeter-scale lifetimes at future beam-dump experiments. In particular, as discussed in \cref{sec:beam-dump}, the SeaQuest experiment can probe much shorter lifetimes than those to which CHARM and NuCal are sensitive. Backgrounds are relatively easy to control for experiments of this type, and they remain sensitive even in our minimal scenario. The figure of merit is the $S$ lifetime, which is at least $\mathcal O(\SI{}{\centi\meter})$ in our minimal scenario. This can be reduced by enhancing the $SP^3$ interaction in our effective theory, but nonetheless, searches for long-lived particles promise to be a powerful probe of our scenario in the coming decade.

\section{Conclusions}
\label{sec:conclusions}

Taken together, the anomalous KOTO events and the Grossman--Nir bound provide a strong hint for light new physics. In this work, we introduced an effective theory that accounts for the excess in the $K_L\to \pi^0 ~ \text{inv.}$ channel with a metastable scalar $S$, a lighter stable pseudoscalar $P$, and effective dimension-five operators that mediate interactions between $S$, $P$ and the $d$ and $s$ quarks. We provided two UV-complete models that would produce an effective theory consistent with our assumptions. We then investigated the implications of our effective theory for cosmology and vice versa. In particular, we showed that cosmological overproduction of $P$ places important constraints on the structure of the low-energy theory.

At face value, in our minimal scenario, $P$ cannot account for either DM or the KOTO excess unless the reheating temperature is close to \SI{10}{\mega\electronvolt}. While it is possible to escape this conclusion by augmenting the model, e.g., with couplings of $P$ to neutrinos, a low reheating temperature is unavoidable in the model's simplest incarnation. However, unless $P$ is very light, the required reheating temperature is compatible with current constraints from BBN and CMB, possibly even offering an observational handle on the model once CMB Stage IV experiments further probe the effective number of relativistic species.

Finally, we discussed three experimental tests of our scenario. First, we showed that portions of our parameter space are within reach of future DM direct detection experiments. Second, our metastable $S$ may be discovered by upcoming long-lived particle searches, particularly the planned SeaQuest upgrade. Finally, if $P$ is in our favored mass range, future KOTO data alone can discriminate between our decay chain and the SM three-body decay on the basis of the neutral pion $p_T$ distribution. There are thus strong discovery prospects for $P$ DM within the next decade.

\begin{acknowledgments}
    The research of W.~A. is supported by the National Science Foundation under Grant No. NSF 1912719. B.~V.~L. and S.~P. are partly supported by the U.S.\ Department of Energy Grant No. de-sc0010107. We thank Maxim Pospelov for introducing us to this class of models for the KOTO excess, and for subsequent discussions. We are grateful to James Unwin for valuable conversations regarding the UV freeze-in paradigm. We thank Stefania Gori for pointing us to relevant beam-dump constraints, and we thank Natalie Telis for valuable conversations concerning statistical methodology.
\end{acknowledgments}

\subsection*{Note Added} 
At the ICHEP 2020 conference, the KOTO collaboration updated the background estimate for their $K_L \to \pi^0 \nu\bar\nu$ search, increasing the number of expected background events to $1.05\pm 0.28$~\cite{KOTO_ICHEP}. This reduces the significance of the observed excess events and shifts the best fit region in figures \ref{fig:Lambda} and \ref{fig:Lambda_SP3} to slightly larger values of $\Lsd$.

\appendix

\section{KOTO simulation} \label{app:MonteCarlo}

In this appendix, we provide details of our calculation of the quantity $R$ introduced in \cref{eq:R}. $R$ is the acceptance of the $K_L \to SP \to \pi^0 PP$ signal relative to the SM $K_L \to \pi^0 \nu\bar\nu$ acceptance at KOTO. Our calculation is based on a Monte Carlo simulation following steps similar to the ones described in~\cite{Kitahara:2019lws,Hostert:2020gou}. 

The layout of the KOTO beam line and the KOTO detector is described e.g. in~\cite{Masuda:2015eta}. We start by generating $K_L$ momenta $p_{K_L}$ and $K_L$ decay vertex locations $z_{K_L}$ based on the distribution
\begin{equation}
    f(p_{K_L},z_{K_L}) \propto g(p_{K_L}) \times \exp\left(
        -\frac{(z_{K_L}-z_\text{exit})m_{K_L}}{\tau_{K_L}p_{K_L}}
    \right) ~,
\end{equation}
where $z_\text{exit} = \SI{20}{\meter}$ is the distance of the beam exit from the target and $g(p_{K_L})$ is the measured $K_L$ momentum distribution at the beam exit from~\cite{Masuda:2015eta}.
We include a small transverse component of the $K_L$ momentum such that the beam profile at the beam exit is constant within an $\SI{8.5}{\centi\meter} \times \SI{8.5}{\centi\meter}$ square and zero outside~\cite{Masuda:2015eta}.

In the case of the SM decay, we generate pion momenta using the $K \to \pi$ form factor from~\cite{Carrasco:2016kpy}. 
In the case of the $K_L \to SP \to \pi^0 PP$ decay, we first generate momenta for $S$, based on the fixed energy of $S$ in the $K_L$ rest frame, $E_S = (m_{K_L}^2 + m_S^2 - m_P^2)/(2m_{K_L})$. We then decay $S$ with a decay length distribution that is determined by the $S \to \pi^0 P$ and $S \to 3P$ partial widths. The pion momentum is generated based on the known pion energy in the $S$ rest frame, $E_{\pi^0} = (m_S^2 + m_{\pi^0}^2 - m_P^2)/(2m_S)$. 

Both in the SM case and the NP case, we let the pion decay promptly into two photons, each with energy $E_\gamma = m_{\pi^0}/2$ in the pion rest frame. We reject events with photons produced less than \SI{2.5}{\meter} after the front face of the front barrel (which starts \SI{1.507}{\meter} after the beam exit), as they would be rejected by photon veto collar counters. All other photons are propagated to the calorimeter located \SI{6.148}{\meter} after the front face of the front barrel~\cite{Masuda:2015eta}. 
The energy and location of the detected photons in the calorimeter are smeared using the parameters given in~\cite{Sato:2015yqa}.

Based on the smeared energy and smeared location of the photons in the calorimeter, the transverse momentum and decay vertex location of the pion is inferred following the procedure described in~\cite{Masuda:2015eta}. If there is more than one solution for the vertex location in the decay volume, we pick the location farther away from the calorimeter. 
We then perform the event selection as in~\cite{Ahn:2018mvc}, taking into account all cuts but timing- and shape-related cuts and the trigger-related cut on the center of energy deposition. We use the updated signal region in the plane of the inferred pion transverse momentum and the pion decay vertex location from~\cite{KOTO}.

The results for $R$ in our benchmark scenarios are shown in \cref{fig:R} as function of the $S$ lifetime.

\bibliography{main}

\end{document}